\documentclass[a4paper,fleqn,usenatbib]{mnras}

\usepackage{newtxtext,newtxmath}

\usepackage[T1]{fontenc}
\usepackage{ae,aecompl}

\usepackage{graphicx}	
\usepackage{amsmath}	
\usepackage{amssymb,pdflscape}	

\title[Reddened Quasar ISM]{The Interstellar Medium Properties of Heavily Reddened Quasars \& Companions at $z\sim2.5$ with ALMA \& JVLA}

\author[M. Banerji et al.]{ \parbox{\textwidth}
{Manda Banerji$^{1,2}$\thanks{E-mail: mbanerji@ast.cam.ac.uk}, 
Gareth C. Jones$^{3,4}$, 
Jeff Wagg$^{5}$, 
Chris L. Carilli$^{3,6}$,
Thomas G. Bisbas$^{7,8,9}$,
Paul C. Hewett$^{1}$
}
\\
$^{1}$Institute of Astronomy, University of Cambridge, Madingley Road, Cambridge CB3 0HA, UK\\
$^{2}$Kavli Institute of Cosmology Cambridge, Madingley Road, Cambridge CB3 0HA, UK\\
$^{3}$National Radio Astronomy Observatory, 1003 Lopezville Road, Socorro, NM 87801, USA\\
$^{4}$Physics Department, New Mexico Institute of Mining and Technology, Socorro, NM 87801, USA \\
$^{5}$SKA Organization, Lower Withington Macclesfield, Cheshire SK11 9DL, UK \\
$^{6}$Battcock Centre for Experimental Astrophysics, Cavendish Laboratory, Cambridge CB3 0HE, UK \\
$^{7}$Department of Astronomy, University of Virginia, Charlottesville, VA 22904, USA \\
$^{8}$Department of Astronomy, University of Florida, Gainesville, FL 32611, USA \\
$^{9}$Max-Planck-Institut f\"ur Extraterrestrische Physik, Giessenbachstrasse 1, D-85748 Garching, Germany
}

\date{Accepted XXX. Received YYY; in original form ZZZ}

\pubyear{2017}

\begin{document}
\label{firstpage}
\pagerange{\pageref{firstpage}--\pageref{lastpage}}
\maketitle

\begin{abstract}
We study the interstellar medium (ISM) properties of three heavily
reddened quasars at $z\sim2.5$ as well as three millimetre-bright companion
galaxies near these quasars.
New JVLA and ALMA observations constrain the CO(1-0),
CO(7-6) and [CI]$^3$P$_2-^3$P$_1$ line emission as well as
the far infrared to radio continuum. The gas excitation and physical properties of the ISM are constrained by comparing our observations to photo-dissociation region
(PDR) models. The ISM in our high-redshift quasars is composed of very
high-density, high-temperature gas which is already highly
enriched in elements like carbon. One of our quasar hosts is shown to be a close-separation ($<$2$\arcsec$)
major merger with different line emission properties in the millimeter-bright galaxy
and quasar components. Low angular resolution observations of high-redshift quasars
used to assess quasar excitation properties should therefore be interpreted with 
caution as they could potentially be averaging over multiple components with different ISM conditions. Our quasars
and their companion galaxies show a range of CO excitation properties spanning the full extent from starburst-like to 
quasar-like spectral line energy distributions. We compare gas masses
based on CO, CI and dust emission, and find that these can disagree when standard assumptions are made regarding
the values of $\alpha_{\rm{CO}}$, the gas-to-dust ratio and the atomic carbon abundances. We conclude that the ISM
properties of our quasars and their companion galaxies are diverse and likely vary spatially across
the full extent of these complex, merging systems.


\end{abstract}

\begin{keywords}
galaxies:evolution -- galaxies:high-redshift -- galaxies:starburst -- galaxies:ISM -- (galaxies):quasars:individual
\end{keywords}



\section{Introduction}

In our conventional picture of massive galaxy formation, the most
luminous starburst galaxies and quasars are assembled via gas-rich
major mergers that trigger both star-formation and accretion onto the
central supermassive black hole (e.g. \citealt{Hopkins:08}), thus
naturally explaining the correlation between black hole mass and
stellar bulge mass in present-day galaxies \citep{Magorrian:98,
Kormendy:13}. Many studies in the literature dating back to
\citet{Sanders:88} have postulated that dusty star-forming galaxies
(DSFGs)/submillimetre galaxies (SMGs), and quasars represent different
evolutionary phases in the merger-driven life-cycle of massive
galaxies, with the same gas supply fuelling the starburst and then
eventually making its way to the central regions in galaxies to fuel
black-hole accretion. Observations of molecular gas in SMGs and
quasars also initially revealed apparent differences between the two
populations. While SMGs were found to have both nuclear, warm gas
components as well as more extended, diffuse colder gas components,
suggestive of a multi-phase molecular interstellar medium (ISM)
\citep{Riechers:11a, Carilli:11}, quasar host galaxies were
generally found to be consistent with a single phase ISM
\citep{Riechers:11b} with the few resolved observations available
suggesting that the gas is distributed on relatively compact scales of
a few kiloparsecs (kpc; \citealt{Carilli:02, Walter:04, Riechers:11b, Willott:13}). In
a simple picture of galaxy formation, these observations could be
explained by quasars representing a later evolutionary stage when the
cold gas reservoirs in the galaxy have been substantially depleted and
the molecular gas in the quasar host galaxy is therefore located in a
compact region close to the central accreting black-hole.
Recently, however, such differences in excitation properties between SMGs/DSFGs
and quasar host galaxies have been called into question
\citep{Sharon:16}, not least because the classification of sources
into separate categories largely stems from historical
definitions and many high-redshift galaxies are undoubtedly hybrids of
both populations. In this paper we focus on a population of hybrid
DSFG/quasar systems in order to shed more light on any putative
evolutionary link between the two populations.

Using wide-field near infra-red surveys, we have discovered a large
population of heavily reddened, hyper-luminous (L$_{\rm{bol}}>10^{47}$
erg/s) quasars that share many of the characteristics of both
DSFGs/SMGs and optically luminous quasars \citep{Banerji:12,
Banerji:13, Banerji:15}. All our reddened quasars have dust
extinctions of A$_{\rm{V}}\sim$2-6 mags, similar to what is observed
in SMGs \citep{Takata:06}, as well as broad emission lines in their
rest-frame optical spectra implying the presence of
10$^{9}$-10$^{10}$M$_\odot$ black-holes, comparable to those in
optically selected quasars. In \citet{Banerji:17} (B17 hereafter), we
presented Atacama Large Millimetre Array (ALMA) observations of the
CO(3-2) molecular line for four of our reddened quasars at redshifts,
$z\sim2-3$. Two further quasars from our sample already had CO
detections in the literature \citep{Feruglio:14, Brusa:15}. All six
quasars were found to have significant reservoirs of molecular gas
providing direct observational evidence that reddened quasars reside
in gas-rich, highly star-forming host galaxies. Our ALMA observations
also revealed the presence of other millimetre-bright, CO-emitting
galaxies near one of the quasars, suggesting that reddened quasars are
associated with over-dense regions in the high-redshift Universe.

In this paper, we present new observations of several more molecular
lines in three of the quasars from B17, as well as their associated
companions, to directly constrain the excitation conditions
and properties of the ISM in these systems. A companion paper (Banerji
et al. 2018b; Paper II hereafter) presents high-resolution observations
of the CO(3-2) line in two of the quasars, to better constrain the gas
dynamics and gas morphologies. The CO(3-2) line
($\nu_{\rm{rest}}$=345.7960 GHz) studied in B17 has a critical density
of n$_{\rm{crit}}\sim$10$^4$cm$^{-3}$, and is therefore tracing warm,
moderately high density gas. By observing both lower
and higher-J rotational transitions of CO such as CO(1-0)
($\nu_{\rm{rest}}$=115.2712 GHz; n$_{\rm{crit}}\sim$10$^3$cm$^{-3}$)
and CO(7-6) ($\nu_{\rm{rest}}$=806.6518 GHz;
n$_{\rm{crit}}\sim$10$^5$cm$^{-3}$), we are able to probe the properties
of the quiescent, cold, gas as well as the very high-density
gas associated with the sites of the most active star formation. The
[CI] $^3$P$_2$-$^3$P$_1$ (CI(2-1) hereafter) line
($\nu_{\rm{rest}}$=809.3420 GHz) lies very close in frequency to
CO(7-6) and can therefore be observed with the same ALMA bandpass
setup as used for the CO(7-6) observations. With a much lower
critical density ($\sim$10$^{3}$cm$^{-3}$), compared to most of the CO
rotational transitions, CI has been argued to be a better tracer of
the total molecular gas mass compared to CO \citep{Papadopoulos:04},
especially in the low-metallicity, high cosmic ray ionization rate
environments, thought to be present in  high-redshift starburst galaxies,
where the CO molecule is more easily destroyed. The new observations
presented in this paper also provide constraints on the [CI] line
luminosities and atomic carbon abundances in our reddened quasars.

The paper is structured as follows. In Section \ref{sec:data} we
present the ALMA and JVLA data that forms the basis for our
analysis. Section \ref{sec:results} summarises the resulting
constraints on the far infrared to radio continuum emission and CO and
[CI] line properties in the sample. In Section \ref{sec:analysis} we
infer dust and gas masses and constrain the ISM properties of our
quasars. Section \ref{sec:conclusions} presents our final
conclusions. Throughout the paper we assume a flat, $\Lambda$CDM
cosmology with H$_0$=70 km s$^{-1}$ Mpc$^{-1}$, $\Omega_{\rm{M}}$=0.3 and
$\Omega_{\Lambda}$=0.7.

\section{DATA}

\label{sec:data}

\subsection{Heavily Reddened Quasars and Companions}

\label{sec:sample}

The sample studied consists of three heavily reddened
quasars from \citet{Banerji:12, Banerji:15} - ULASJ0123$+$1525
($z_{H\alpha}=2.630$), ULASJ1234$+$0907 ($z_{H\alpha}=2.503$) and ULASJ2315$+$0143
($z_{H\alpha}=2.561$) - where CO(3-2) emission has already been detected (B17). 
These quasars are among the reddest sources that have been 
discovered in our near infra-red search for a population of heavily 
reddened, intrinsically luminous quasars that are not present in wide-field 
optical surveys such as the Sloan Digital Sky Survey (SDSS). The three quasars
studied here have $(H-K)$ colours of $\gtrsim$2 (Vega), which corresponds to typical
dust extinctions towards the quasar continuum of E(B-V)$\gtrsim$1 (A$_{\rm{V}}\gtrsim3$ mag for R$_{\rm{V}}$=3.1).
In B17, we also detected CO(3-2) emission associated with
two neighbouring galaxies near ULASJ1234$+$0907 - G1234N
($\sim$21$\arcsec$ from quasar; $z_{\rm{CO}}=2.514$) and G1234S
($\sim$11.5$\arcsec$ from quasar $z_{\rm{CO}}=2.497$), which are also studied
further in this paper. Although both galaxies are also covered by the UKIDSS-LAS survey, from which the quasar
was initially selected, they are too faint at near infra-red wavelengths to be detected in UKIDSS. Both galaxies are
likely to be highly obscured and our ALMA observations
in B17 therefore correspond to the first detections of these new galaxies.
We begin by describing the different
observations conducted for each of these sources.

\subsection{Karl G. Jansky Very Large Array (JVLA) Ka-band Observations}

\label{sec:eVLA}

We observed all three quasars with the JVLA in the Ka-band, with the
aim of detecting any cold gas reservoirs as traced by the CO(1-0)
transition at $\nu_{\rm{rest}}$=115.27120 GHz. The JVLA primary beam
also encompasses the two companion galaxies to ULASJ1234$+$0907 at these frequencies, so
constraints on their CO(1-0) line properties were also
obtained. Observations were conducted in 2017 February - March using
the compact D configuration. The correlator was configured to 64 bands,
each with a bandwidth of 128 MHz. The channel width is 2 MHz. Data
were calibrated in the first instance using the VLA Common Astronomy
Software Applications package \textsc{casa} v4.7.1. We found
that some additional flagging of visibilities was required in the case
of both ULASJ1234 and ULASJ2315 before imaging. Spectral line cubes as
well as continuum images were produced using the \textit{clean}
algorithm, employing a natural weighting for the visibilities. In
the case of the spectral line data, the continuum was subtracted in
the u-v plane using two spectral windows either side of the line as
well as the line-free channels in the spectral window containing the
line. The reduction
pipeline applies Hanning-smoothing to the data by default. More details of these observations can be found in Table
\ref{tab:vla_obs}.

The brighter of the two companion galaxies to ULASJ1234 - G1234N, was
detected in CO(1-0) emission with the JVLA although the fainter galaxy, G1234S was not. Line 
images were constructed from the primary-beam corrected \textit{clean} images centred on the 
quasar. No continuum emission has been detected from
either companion galaxy over the entire bandwidth of the JVLA Ka-band
and we therefore did not perform continuum subtraction. 

\begin{table*}
\begin{center}
\caption{Summary of JVLA Ka-band observations}
\label{tab:vla_obs}
\begin{tabular}{lccc}
& ULASJ0123$+$1525 & ULASJ1234$+$0907 & ULASJ2315$+$0143 \\
\hline
Dates Observed & 2017-03-13 & 2017-02-11,2017-02-12,2017-02-17 & 2017-03-03, 2017-03-04 \\
Number of Antennae & 27 & 27 & 27 \\
Exposure Time & 5h & 15h & 10h \\
Beam Size (line) / $\arcsec$ & 2.81$\times$2.32 & 2.46$\times$2.36 & 2.98$\times$2.07 \\
Beam size (continuum) / $\arcsec$ & 2.77$\times$2.35 & 2.51$\times$2.38 & 2.66$\times$2.11 \\
Channel R.M.S in 2MHz channels / mJy beam$^{-1}$  &  0.10 & 0.06 &  0.10 \\
Continuum R.M.S / $\mu$Jy beam$^{-1}$ & 3.5 & 2.2 & 5.1 \\
\hline 
\end{tabular}
\end{center}
\end{table*}

\subsection{ALMA Band 6 Observations}

\label{sec:alma_band6}

Two of the quasars - ULASJ1234 and ULASJ2315 - were also
observed using ALMA Band 6 with the aim of probing the high-excitation
gas via the CO(7-6) transition, as well as the dust continuum at
observed frame wavelengths of $\sim$1.2mm. Observations were conducted
in 2017 March using the most compact available configuration. The
correlator was configured to four dual polarization bands of 2 GHz
(1.875 GHz effective) bandwidth each providing a channel width of
15.625 MHz. Two of the basebands were centred on the CO(7-6) and
CI(2-1) transitions respectively, while the other two basebands were
positioned so as to obtain a measurement of the dust continuum
emission. More details of these observations can be found in Table
\ref{tab:obs_alma}.

\begin{table*}
\begin{center}
\caption{Summary of ALMA Band 6 Observations}
\label{tab:obs_alma}
\begin{tabular}{lcc}
& ULASJ1234$+$0907 & ULASJ2315$+$0143 \\
\hline
Date Observed  & 2017-03-25 & 2017-03-25 \\
Number of Antennae & 43 & 41 \\
Exposure Time / mins & 36.8 & 52.3 \\
Beam Size (line) / $\arcsec$ & 1.96$\times$1.76 & 1.98$\times$1.72 \\
Beam Size (continuum) / $\arcsec$ & 1.84$\times$1.66 & 1.66$\times$1.42 \\
Channel R.M.S in 15.6 MHz channels / mJy beam$^{-1}$  & 0.27 & 0.24 \\
Continuum R.M.S / $\mu$Jy beam$^{-1}$ & 39 & 41 \\
\hline 
\end{tabular}
\end{center}
\end{table*}

All data were calibrated and reduced using the ALMA pipeline in
\textsc{casa} (v4.7.0) by executing the appropriate ALMA calibration
scripts corresponding to the release date of the observations. Time
dependent amplitude and phase variations were calibrated using nearby
quasars while flux calibrations made use of observations of Ganymede
and Neptune. Dust continuum images and spectral line cubes were produced
from the calibrated visibilities using the \textsc{casa} task
\textit{clean}. In the case of ULASJ1234 we used natural weighting of
the visibilities to produce both the line and continuum images. For
ULASJ2315, natural weighting was used to produce the line image but a
Briggs weighting with \textit{robust}=0.5 was found to produce a
better compromise between sensitivity and resolution when producing
the continuum image. In B17 we serendipitously detected CO(3-2)
emission from two companion galaxies to ULASJ1234. The galaxy closest
to the quasar - G1234S - lies within the full-width-half-maximum of
the ALMA primary beam in Band 6, but the second companion galaxy -
G1234N - does not. We constructed continuum and line images for G1234S
from the primary-beam corrected image centred on the quasar, although
we note that G1234S lies very close to the 10\% response level of the beam where
\textsc{casa} truncates the image by default. A natural weighting of the visibilities was used. 

For the spectral line cubes, continuum subtraction was carried out in
the u-v plane using the \textsc{casa} task \textit{uvcontsub},
employing the line-free channels to determine the continuum level. The pipeline 
reduction applies Hanning smoothing to the spectra by default. The
final beam sizes for both the line and continuum images, the typical
root mean square (RMS) sensitivities, as well as other details of the
observations, are summarised in Table \ref{tab:obs_alma}.

\subsection{ALMA Band 3 Observations}

ALMA Band 3 observations tracing CO(3-2) emission as well as continuum
emission at rest-frame frequencies of $\sim$300-370 GHz were presented
for all three quasars in B17. In Paper II we present new higher angular
resolution, higher signal-to-noise ratio (S/N) observations of ULASJ2315 and
ULASJ1234, which represent a factor of $\sim$3 improvement in spatial resolution 
(from $\sim$3$\arcsec$ to $<$1$\arcsec$) compared to the observations in B17.
Both the continuum flux densities and CO(3-2) line
properties inferred from these observations are used later in this
paper. The results from B17 and Paper II are therefore 
summarised in Table \ref{tab:all_obs} but readers should refer to these papers for further details of these measurements. Whenever possible, we make use
of the higher S/N measurements from Paper II. In Paper II we
show that the CO(3-2) emission from ULASJ2315 is clearly resolved
both spatially and spectrally into two distinct components with $\Delta$r$\sim$1.9$\arcsec$(=15 kpc at $z=2.566$) and $\Delta$v$\sim$160 km/s, and this system is therefore a close
separation merger. In several of the observations presented here, we
do not have the necessary angular resolution to be able to separate
the emission from the two components of the merger. We therefore
consider both integrated properties of the system (from the low
angular resolution data in B17) as well as spatially resolved
properties of the quasar host galaxy and the nearby companion galaxy
(presented in detail in Paper II). Readers are referred to Paper II for details of
the dynamical analysis of the gas emission in both ULASJ2315 and
ULASJ1234 and their associated companion galaxies including a 
comparison of the CO(3-2) gas dynamics with the kinematics of some of the other molecular lines studied in this work.

\subsection{Line and Continuum Measurements}

In the following section we use our new data to derive constraints on the continuum and emission line fluxes in our
quasar host galaxies and companions. In all cases continuum flux densities are estimated by fitting a 2-dimensional 
elliptical Gaussian to the continuum source in the image plane, although we have checked that our results do not change
if the fitting is done in the u-v plane instead. In the case of the emission lines, we first extract 1-dimensional spectra from the
continuum subtracted line cubes and then derive line properties by fitting either a single or double Gaussian to the emission 
line in the 1-d spectrum as detailed below.

\section{Results}

\label{sec:results}

\begin{landscape}
\begin{table*}
\begin{center}
\caption{Observed continuum and line properties of our heavily reddened quasars and their companion galaxies. Upper limits are quoted at the 3$\sigma$ level and are appropriate for a point source.}
\label{tab:all_obs}
\begin{tabular}{lccccccc}
& ULASJ0123 & ULASJ1234 & G1234N & G1234S & ULASJ2315 (Tot) & ULASJ2315 (QSO) & ULASJ2315 (Gal) \\
\hline
CO Redshift from B17 or Paper II & 2.630 & 2.503 & 2.514 & 2.497 & 2.566 & 2.566 & 2.566 \\
\hline
\multicolumn{8}{c}{\bf JVLA Ka-band Observations (This Paper)} \\ 
Continuum $\nu_{\rm{obs}}$ / GHz & 33.1730 & 33.7492 & 33.7492 & 33.7492 & 33.4775 & -- & -- \\
Continuum S$_\nu$ / $\mu$Jy beam$^{-1}$ & 27$\pm$2 & 57$\pm$2$^{\dagger}$ & $<$6.6 & $<$6.6 & 350$\pm$20$^{\dagger}$ & -- & -- \\
CO(1-0) $\Delta$FWHM / km s$^{-1}$ & 450$\pm$60 & 1100$\pm$470 & 450$\pm$240 & 530$^{\ast}$ & 450$\pm$210 & -- & -- \\
CO(1-0) Line Intensity / Jy km s$^{-1}$ & 0.11$\pm$0.01 & 0.11$\pm$0.01 & 0.07$\pm$0.01 & $<$0.05 & 0.08$\pm$0.01 & -- & -- \\
CO(1-0) Line Luminosity / $\times$10$^6$ L$_\odot$ & 1.7$\pm$0.2 & 1.6$\pm$0.2 & 1.1$\pm$0.1 & $<$0.70 & 1.1$\pm$0.2 & -- & -- \\
L'$_{\rm{CO(1-0)}}$ / $\times$10$^{10}$ K km s$^{-1}$ pc$^2$ & 3.7$\pm$0.4 & 3.6$\pm$0.4 & 2.3$\pm$0.3 & $<$1.6 & 2.6$\pm$0.4 & -- & -- \\
\hline
\multicolumn{8}{c}{\bf ALMA Band 6 Observations (This Paper)} \\ 
Continuum $\nu_{\rm{obs}}$ / GHz  & -- & 245.8170 & -- & 245.8170 & 242.0485 & 242.0485 & 242.0485 \\
Continuum S$_\nu$ / mJy beam$^{-1}$ & -- & 1.33$\pm$0.03 & -- & 0.91$\pm$0.04 & 0.92$\pm$0.14$^{\dagger}$ & 0.42$\pm$0.04 & 0.44$\pm$0.12$^{\dagger}$ \\
CO(7-6) $\Delta$FWHM / km s$^{-1}$ & -- & 900$\pm$80 & -- & 920$\pm$330 & 700$\pm$180 & 580$\pm$70 & 200$\pm$70 \\
CO(7-6) Line Intensity / Jy km s$^{-1}$ & -- & 1.97$\pm$0.07 & -- & 1.06$\pm$0.06 & 0.52$\pm$0.07 & 0.45$\pm$0.10 & 0.10$\pm$0.03 \\
CO(7-6) Line Luminosity / $\times$10$^8$ L$_\odot$ & -- & 1.96$\pm$0.07 & -- & 1.06$\pm$0.06 & 0.54$\pm$0.07 & 0.47$\pm$0.10 & 0.11$\pm$0.04 \\
L'$_{\rm{CO(7-6)}}$ / $\times$10$^{10}$ K km s$^{-1}$ pc$^2$ & -- & 1.25$\pm$0.05 & -- & 0.67$\pm$0.04 & 0.35$\pm$0.05 & 0.30$\pm$0.07 & 0.07$\pm$0.02 \\
CI(2-1) $\Delta$FWHM / km s$^{-1}$ & -- & 630$\pm$90 & -- & 880$\pm$320 & 520$\pm$120 & 620$\pm$90 & 60$\pm$50 \\
CI(2-1) Line Intensity / Jy km s$^{-1}$ & -- & 0.59$\pm$0.07 & -- & 1.09$\pm$0.06 & 0.49$\pm$0.07 & 0.47$\pm$0.10 & 0.10$\pm$0.03 \\
CI(2-1) Line Luminosity / $\times$10$^7$ L$_\odot$ & -- & 5.8$\pm$0.7 & -- & 10.8$\pm$0.6 & 5.2$\pm$0.7 & 4.9$\pm$1.0 & 1.0$\pm$0.4 \\
L'$_{\rm{CI(2-1)}}$ / $\times$10$^{9}$ K km s$^{-1}$ pc$^2$ & -- & 3.7$\pm$0.5 & -- & 6.8$\pm$0.4 & 3.2$\pm$0.5 & 3.1$\pm$0.7 & 0.6$\pm$0.2 \\
\hline
\multicolumn{8}{c}{\bf ALMA Band 3 Observations (B17 and Paper II)} \\ 
Continuum $\nu_{\rm{obs}}$ / GHz & 100.8415 & 86.2308 & 91.4662 & 91.4662 & 102.6883 & 102.6042 & 102.6042 \\
Continuum S$_\nu$ / $\mu$Jy beam$^{-1}$ & 98$\pm$16 & 71$\pm$7 & 38$\pm$5 & 34$\pm$2 & 259$\pm$17$^{\dagger}$ & 160$\pm$8 & 45$\pm$7$^{\dagger}$ \\
CO(3-2) $\Delta$FWHM / km s$^{-1}$ & 520$\pm$40 & 870$\pm$60 & 560$\pm$70 & 530$\pm$80 & 350$\pm$50 & 300$\pm$60 & 190$\pm$40 \\
CO(3-2) Line Intensity / Jy km s$^{-1}$ & 1.40$\pm$0.08 & 0.93$\pm$0.02 & 0.80$\pm$0.02 & 0.45$\pm$0.02 & 0.91$\pm$0.05 & 0.27$\pm$0.02 & 0.28$\pm$0.01 \\
CO(3-2) Line Luminosity / $\times$10$^7$ L$_\odot$ & 6.5$\pm$0.3 & 3.97$\pm$0.09 & 3.40$\pm$0.07 & 1.93$\pm$0.07 & 4.0$\pm$0.2 & 1.19$\pm$0.07 & 1.26$\pm$0.06 \\
L'$_{\rm{CO(3-2)}}$ / $\times$10$^{10}$ K km s$^{-1}$ pc$^2$ & 5.3$\pm$0.3 & 3.21$\pm$0.07 & 2.77$\pm$0.06 & 1.56$\pm$0.05 & 3.2$\pm$0.2 & 0.96$\pm$0.06 & 1.01$\pm$0.05 \\
\hline
\end{tabular}
\flushleft
\small{$^\ast$ Fixed to $\Delta$FWHM of CO(3-2) line from Paper II} \\
\small{$^\dagger$ Spatially extended over multiple beams. Integrated emission over entire spatial extent (units of $\mu$Jy or mJy)} \\
\end{center}
\end{table*}
\end{landscape}

\subsection{Far Infra-red to Radio Continuum}

\label{sec:cont_obs}

\begin{figure*}
\begin{center}
\begin{tabular}{ccc}
\includegraphics[scale=0.45]{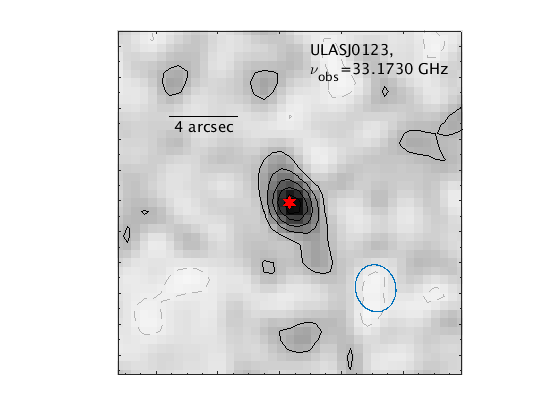} & \hspace{-1.5cm} \includegraphics[scale=0.45]{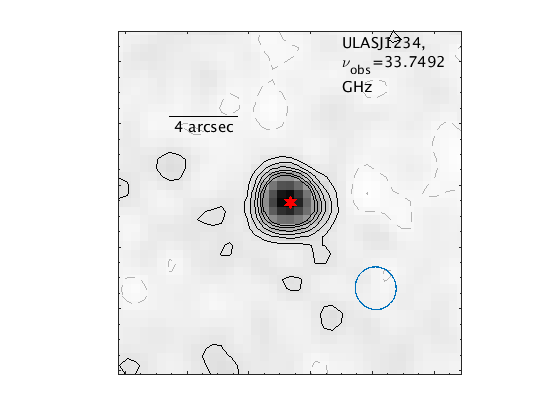} & \hspace{-1.5cm} \includegraphics[scale=0.45]{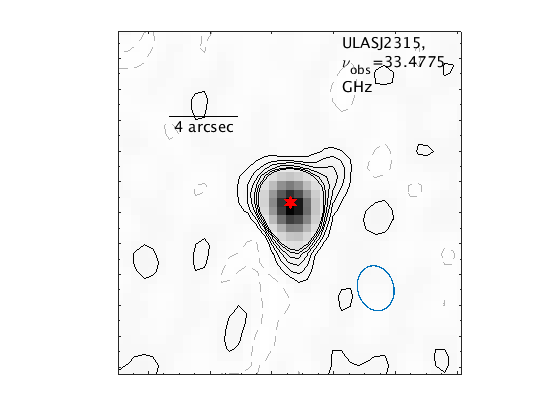} \\
\end{tabular}
\caption{20$\times$20$\arcsec$ continuum images at $\nu_{\rm{obs}}\sim$33 GHz produced from the JVLA data for ULASJ0123 (left), ULASJ1234 (middle) and ULASJ2315 (right). The solid and dashed lines represent positive and negative flux contours at intervals of 1.5$\sigma$. The location of the quasar is marked with a star and the beam size and position angle is indicated in the bottom right-corner of each image. North is up and East is to the right.}
\label{fig:vlacont}
\end{center}
\end{figure*}

All of our quasars observed with the JVLA were detected in the
continuum at rest-frame frequencies of $\sim$118-120 GHz. The
continuum images are shown in Fig. \ref{fig:vlacont} and the flux
densities are summarised in Table \ref{tab:all_obs}. Even at the low
angular resolution of the JVLA data (major axis of beam
$\sim$2.5$\arcsec$), two of the three quasars are spatially resolved
in continuum emission. ULASJ1234 has a spatial extent of
$\sim$1.3$\arcsec$ while ULASJ2315 has a spatial extent of
$\sim$0.8$\arcsec$ along the major axis (after deconvolving from the
beam). Neither of the companion galaxies to ULASJ1234 are detected in
the continuum and 3$\sigma$ upper limits on their
flux densities are quoted in Table \ref{tab:all_obs}.

\begin{figure*}
\begin{center}
\begin{tabular}{ccc}
\includegraphics[scale=0.45]{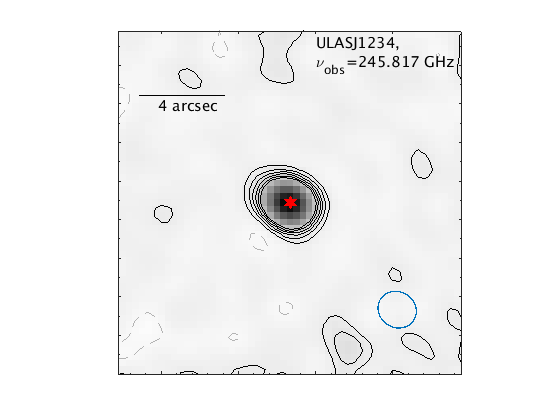} & \hspace{-1.5cm} \includegraphics[scale=0.45]{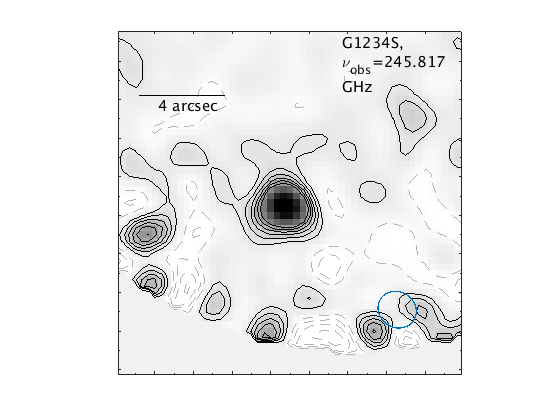} & \hspace{-1.5cm} \includegraphics[scale=0.45]{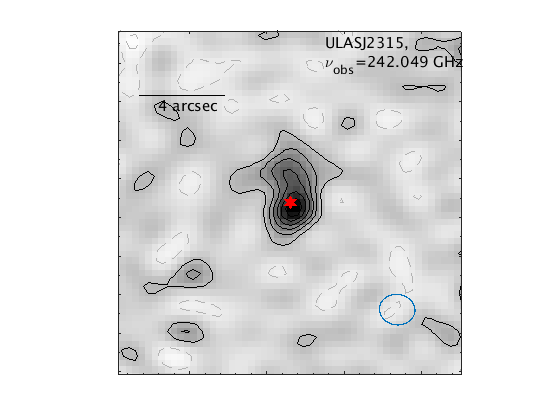} \\
\end{tabular}
\caption{16$\times$16$\arcsec$ continuum images at $\nu_{\rm{obs}}\sim$245 GHz produced from the ALMA Band 6 data for ULASJ1234 (left), G1234S (middle) and ULASJ2315 (right). The solid and dashed lines represent positive and negative flux contours at intervals of 1.5$\sigma$. The location of the quasars is marked with a star and the beam size and position angle is indicated in the bottom right-corner of each image. North is up and East is to the right.}
\label{fig:dustcont}
\end{center}
\end{figure*}

We also detected continuum emission at rest-frame frequencies of
$\sim$859-862 GHz from both quasars observed with ALMA in Band 6 as well as the
single companion galaxy to ULASJ1234 that falls within the ALMA Band 6
primary beam. The continuum flux densities are again summarised in
Table \ref{tab:all_obs} and the images are shown in
Fig. \ref{fig:dustcont}. The continuum emission is unresolved, or only
marginally resolved, in the case of ULASJ1234 and its companion galaxy
G1234S. However, in the case of the close-separation merger ULASJ2315
(Paper II), spatially extended continuum emission is seen, with the two
components of the merger evident in the image in
Fig. \ref{fig:dustcont}. Using the \textsc{casa} task \textit{imfit},
we model the dust continuum image as the sum of two 2-dimensional
Gaussian components. One of the Gaussian components is centred near
the quasar while the second has a best-fit centroid that is
$\sim$1.85$\arcsec$ ($\sim$15 kpc) to the north of the quasar. The
distance is consistent with the physical separation between the two
components of the merger inferred from the CO(3-2) gas emission in
Paper II. The dust emission from the quasar is unresolved but the
emission from the northern companion galaxy is extended with a
deconvolved size of $\sim$2.0$\arcsec$ along the major axis as
inferred from \textit{imfit}. The continuum flux densities of both
components of the merger are also presented in Table
\ref{tab:all_obs}. From hereon, we refer to these two components of
the merger as the `QSO' and `GAL' components of ULASJ2315.

\subsection{CO(1-0) Line Properties}

\label{sec:CO10}

\begin{figure*}
\begin{center}
\begin{tabular}{cc}
\includegraphics[scale=0.45]{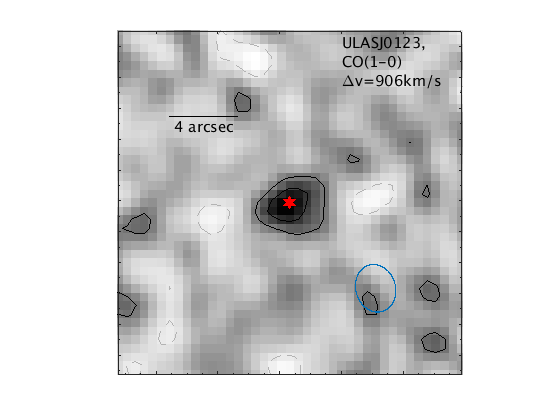} & \hspace{-1cm} \includegraphics[scale=0.45]{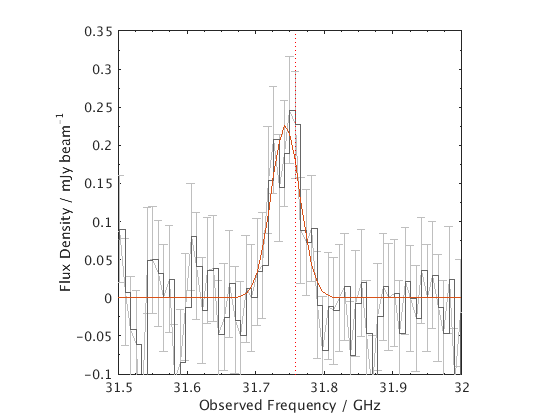} \\
\includegraphics[scale=0.45]{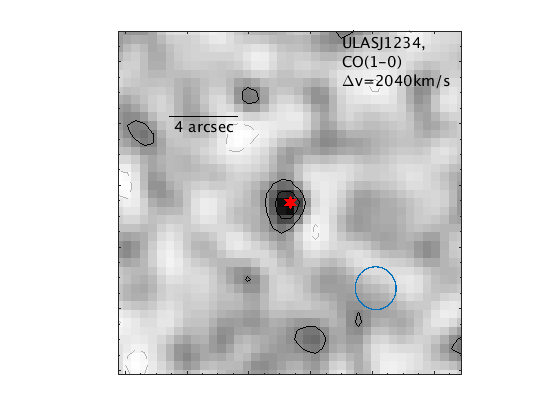} & \hspace{-1cm} \includegraphics[scale=0.45]{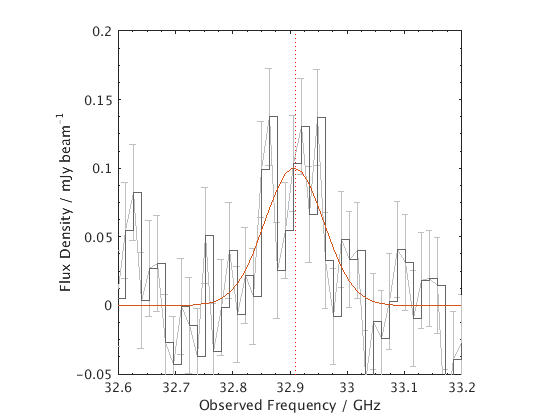} \\
\includegraphics[scale=0.45]{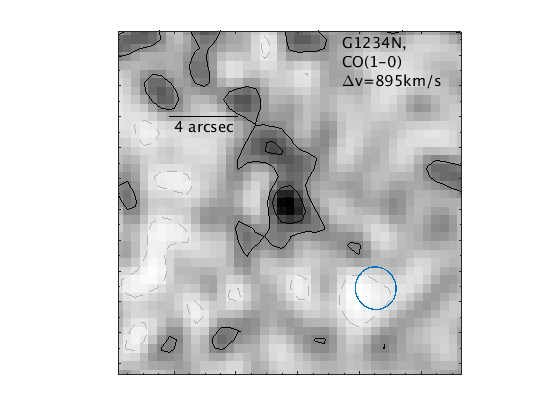} & \hspace{-1cm} \includegraphics[scale=0.45]{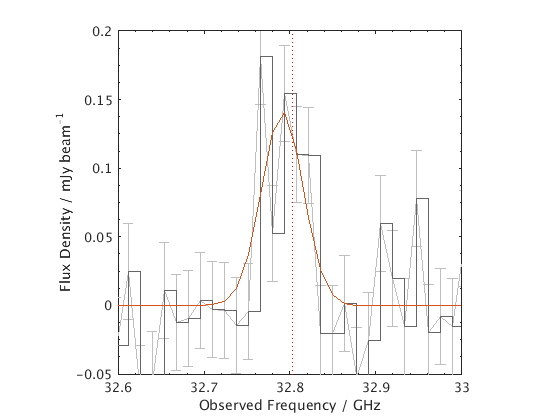} \\
\includegraphics[scale=0.45]{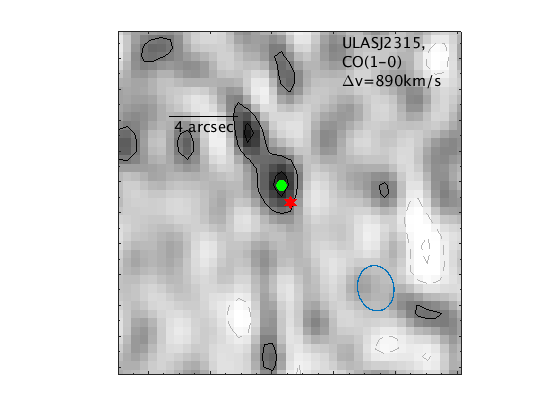} & \hspace{-1cm} \includegraphics[scale=0.45]{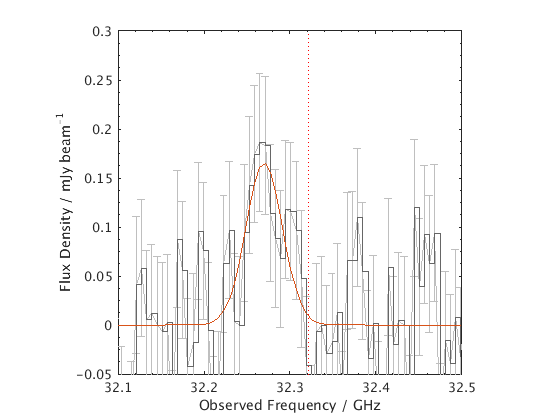} \\
\end{tabular}
\caption{\textit{Left:} 20$\times$20$\arcsec$ CO(1-0) line maps integrated over $\pm$FWHM of the line profile corresponding to velocity intervals of $\Delta$v as indicated in each panel. The solid and dashed lines represent positive and negative flux contours at intervals of 1.5$\sigma$. The location of the quasars is marked with a star and the beam size and position angle are indicated in the bottom right-corner of each image. North is up and East is to the right. \textit{Right:} CO(1-0) spectral line profiles together with the best-fit single Gaussian model. Error-bars denote the 1$\sigma$ RMS per channel. Vertical dotted lines mark the expected position of CO(1-0) based on the CO(3-2) derived redshifts for the sources in B17 and Paper II. For ULASJ2315 (bottom panels), the peak of the CO(1-0) emission is spatially offset from the quasar centroid and the CO(1-0) spectrum has therefore been extracted from the position of the green circle. The CO(1-0) line in ULASJ2315 is also shifted in velocity relative to CO(3-2).}
\label{fig:CO10_spec}
\end{center}
\end{figure*}

We detected CO(1-0) emission from the three quasars observed with the
JVLA as well as from the brighter companion galaxy to the quasar
ULASJ1234 - G1234N. The S/N of all the CO(1-0) detections are
relatively modest at the native resolution of the data. We have,
therefore, re-binned the spectral axis by factors of 4, 7, 7 and 4 for
ULASJ0123, ULASJ1234, G1234N and ULASJ2315 respectively, to provide
five or six resolution elements across the the full width at half
maximum (FWHM) of the line, as inferred from the CO(3-2) observations
in B17 and Paper II. The CO(1-0) maps together with the binned CO(1-0)
spectra for the four sources are shown in Fig. \ref{fig:CO10_spec}.

We model the CO(1-0) line profiles using a single Gaussian and derive
line properties (see Table \ref{tab:all_obs}) from the model. The
second companion galaxy near the quasar ULASJ1234 - G1234S - is
undetected in CO(1-0) and we derive an upper limit on the CO(1-0)
brightness temperature, assuming that the line has the same width and
velocity centroid as the CO(3-2) emission from this source, which is
detected at high S/N in Paper II.

For ULASJ2315, the peak of the CO(1-0) emission appears to be
spatially offset from the quasar by 1.1$\pm$0.5$\arcsec$. No such offset is seen in the JVLA
Ka-band continuum image in Fig. \ref{fig:vlacont} suggesting that the
offset between the quasar and the CO(1-0) emission is likely real. In
Paper II, and in the ALMA dust continuum image shown in
Fig. \ref{fig:dustcont}, we have seen that ULASJ2315 is a merger of
two galaxies. Although our JVLA observations lack the spatial
resolution to be able to resolve the two components, the peak in the
CO(1-0) emission is spatially coincident with the northern companion
galaxy rather than the quasar. The spectrum in
Fig. \ref{fig:CO10_spec} is extracted at the peak position of the
CO(1-0) emission rather than at the quasar centroid, and also shows a
velocity offset of 500$\pm$100 km s$^{-1}$ relative to the CO(3-2) emission from the quasar. This
may also be suggestive of different kinematics affecting the cold and warm gas
reservoirs in this system although higher resolution, higher S/N data would be needed to confirm this.

\subsection{CO(7-6) and CI(2-1) Line Properties}

\label{sec:CO_CI}

\begin{figure*}
\begin{center}
\begin{tabular}{cc}
\includegraphics[scale=0.45]{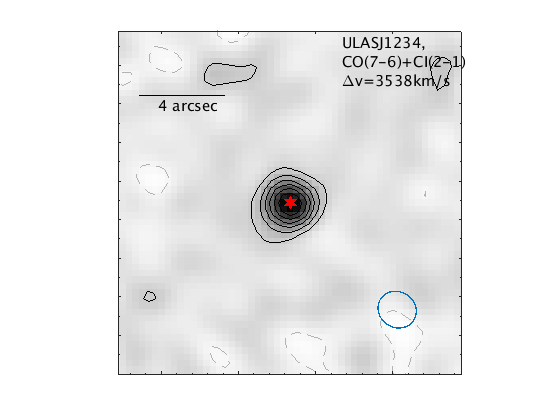} & \hspace{-1cm} \includegraphics[scale=0.45]{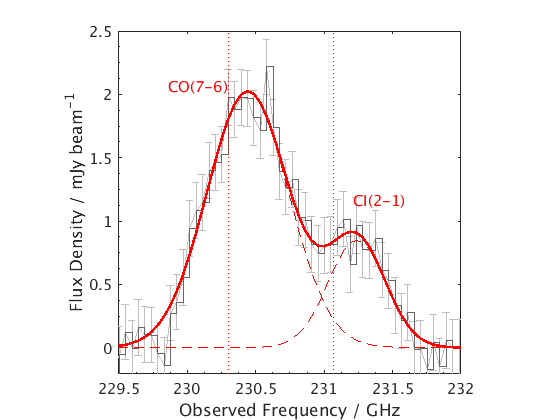} \\
\includegraphics[scale=0.45]{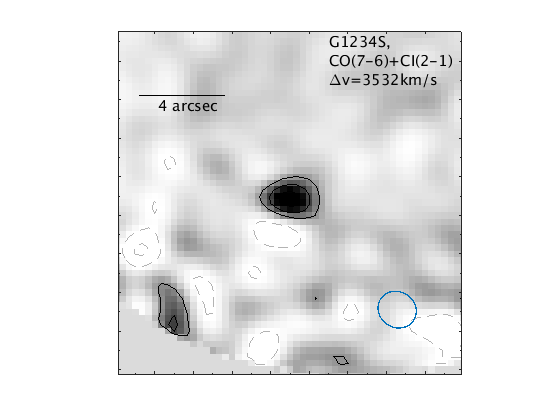} & \hspace{-1cm} \includegraphics[scale=0.45]{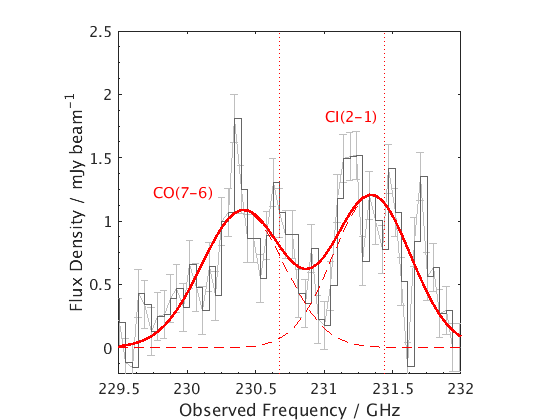} \\
\end{tabular}
\caption{\textit{Left:} 16$\times$16$\arcsec$ CO(7-6)+CI(2-1) line maps integrated over the two subbands containing both lines for ULASJ1234 (top) and its companion galaxy G1234S (bottom). The solid and dashed lines represent positive and negative flux contours at intervals of 1.5$\sigma$. The location of the quasar is marked with a star and the beam size and position angle is indicated in the bottom right-corner of each image. North is up and East is to the right. \textit{Right:} CO(7-6) and CI(2-1) spectral line profiles together with the best-fit model to these line profiles represented by the sum (solid) of two (dashed) Gaussian profiles. Error-bars denote the 1$\sigma$ RMS per channel. Vertical dotted lines mark the expected position of these lines based on the CO(3-2) derived redshifts for the sources in B17 and Paper II. }
\label{fig:CO76_CI_spec_1234}
\end{center}
\end{figure*}

\begin{figure*}
\begin{center}
\begin{tabular}{cc}
\includegraphics[scale=0.45]{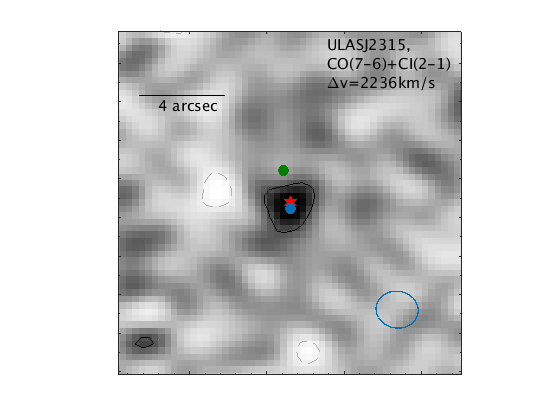} & \hspace{-1cm} \includegraphics[scale=0.45]{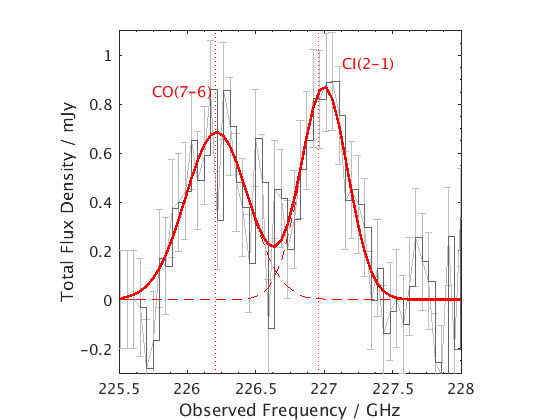} \\
\includegraphics[scale=0.45]{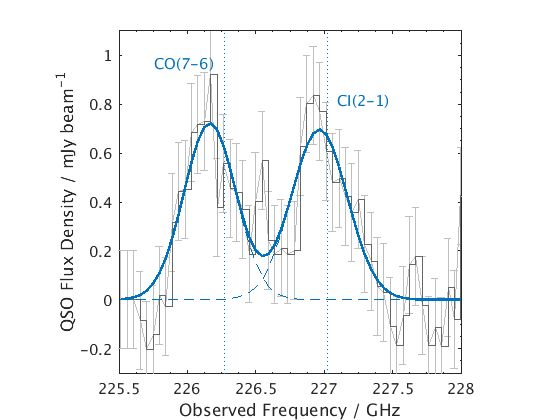} & \hspace{-1cm} \includegraphics[scale=0.45]{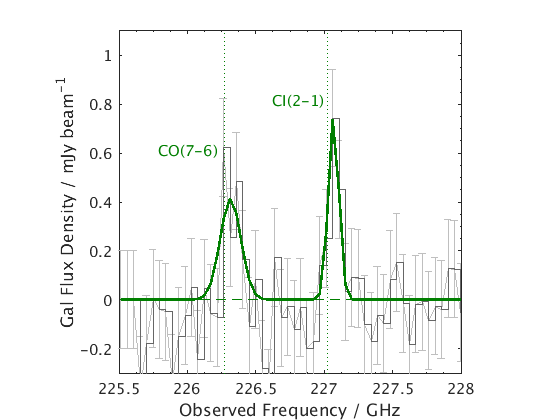} \\
\end{tabular}
\caption{\textit{Top Left:} 16$\times$16$\arcsec$ CO(7-6)+CI(2-1) line maps integrated over $\pm$FWHM of both lines in the ULASJ2315 system corresponding to a velocity interval $\Delta$v=2236 km/s. The solid and dashed lines represent positive and negative flux contours at intervals of 1.5$\sigma$. The location of the quasar is marked with a star while the blue and green circles mark the centroids of the two components corresponding to the quasar host and the northern companion inferred from the high angular resolution observations of the CO(3-2) emission in Paper II. North is up and East is to the right. In the top-right we show the spectrum integrated over the entire spatial extent of the source while the bottom two panels show the spectra extracted from the centroids of the two components of the merger. The best-fit model to these line profiles is the sum (solid) of two (dashed) Gaussian profiles. Error-bars denote the 1$\sigma$ RMS per channel. Vertical dotted lines mark the expected positions of these lines based on the CO(3-2) derived redshifts for the sources in Paper II. }
\label{fig:CO76_CI_spec_2315}
\end{center}
\end{figure*}

CO(7-6) and CI(2-1) line emission is detected from the two
quasars targeted with ALMA as well as the companion galaxy G1234S. 
The spectra covering both lines, as well as the line
maps in this region, can be seen in Fig. \ref{fig:CO76_CI_spec_1234}
and Fig. \ref{fig:CO76_CI_spec_2315}. We model the combined CO(7-6) and
CI(2-1) line profiles as the sum of two Gaussian components. In the
case of ULASJ2315 we consider the integrated line emission from both
galaxies in the system, as well as the ``resolved" line emission. The 
latter is estimated by assuming that the CO(7-6) and CI(2-1) emission is
unresolved in each component of the merger in the current Band 6 data, 
but spatially co-incident with the CO(3-2) emission, which is clearly resolved in Paper II.
The derived line properties for all sources are summarised in Table
\ref{tab:all_obs}.

As can be seen in Fig. \ref{fig:CO76_CI_spec_2315} and Table
\ref{tab:all_obs}, the line properties change markedly across the two
components of the merger in the ULASJ2315 system. In particular, the
quasar host galaxy appears to have broad emission lines whereas the companion northern
galaxy is characterised by much narrower line emission and a higher CI(2-1)
peak flux-density compared to CO(7-6). The CO(7-6) lines are generally broader than the CI(2-1) lines
but the difference is only statistically significant in the case of
ULASJ1234 and the galaxy component of ULASJ2315. In both cases, the
difference suggests that the high-density, highly excited gas traced
by CO(7-6) is probably coming from a different region of the galaxy
compared to the lower-density gas traced by CI(2-1).

\section{Analysis}

\label{sec:analysis}

\subsection{Dust Continuum \& Non-Thermal Spectral Energy Distribution}

\label{sec:cont}

The new ALMA Band 6 and JVLA Ka-band continuum observations when
combined with our ALMA Band 3 observations (B17 and Paper
II), now allow us to sample the radio to far infra-red spectral energy
distribution in these sources \citep{Condon:92}. The extended
frequency range allows emission from the Rayleigh-Jeans tail of the
dust spectral energy distribution (SED) to potentially be separated from free-free
emission and non-thermal synchrotron emission. In the case of both
ULASJ1234 and ULASJ2315, we sample the far infra-red to radio
continuum at $\nu_{\rm{obs}}\sim$ 240GHz, 100GHz and 33GHz
($\nu_{\rm{rest}} \sim$840-850 GHz, 350-360GHz, 115-117GHz). In the 
case of ULASJ0123 we have two continuum data points at  $\nu_{\rm{obs}}\sim$ 100 GHz and 33 GHz. 
ULASJ1234 was also detected using \textit{Herschel} \citep{Banerji:14}
although due to the poorer angular resolution of \textit{Herschel} the photometry
is a blend of the quasar and the two other companion galaxies located near it
(Section \ref{sec:sample}). ULASJ1234 and ULASJ2315 are both covered by the VLA-FIRST
survey and both quasars are undetected at $\nu_{\rm{obs}}$=1.4 GHz with flux density limits of $<$1 mJy and $<$0.93 mJy
respectively.

We fit SED models that are a composite of a
power-law synchrotron component, which dominates at frequencies below
$\sim$30 GHz, a thermal free-free component with a power-law index of
$\alpha_{\rm{ff}}=-0.1$ and a single temperature greybody component
described by a dust temperature, T$_{\rm{d}}$ and emissivity index,
$\beta$.  Due to the limited number of photometric points we fix the
parameters of the greybody to T$_{\rm{d}}$=47K and $\beta$=1.6
\citep{Beelen:06}, although the constraints are similar
for $\beta$=1.6$-$2.4. For ULASJ2315 we consider integrated fluxes
from the entire merger system as we do not resolve the individual components of the merger at
all frequencies. The SED-fit parameters for this system should
therefore be interpreted with caution as discussed in detail below.

In line with previous studies (e.g. \citealt{Wagg:14}), we characterize
the SED of each galaxy by assuming the linear relationships from
\citet{Yun:02} between the total star formation rate and the three
components describing the radio to far infra-red spectral energy
distribution - the non-thermal synchrotron ($S_{\rm{nth}}$), free-free ($S_{\rm{ff}}$), and thermal dust ($S_{\rm{d}}$) emission:

\begin{equation}
S_{\rm{nth}}(\nu)=25 f_{\rm{nth}} \nu^{-\alpha} \frac{\rm{SFR}}{M_\odot \rm{yr}^{-1}} {D_{\rm{L}}}^{-2} \rm{Jy}
\label{eq:ssynch}
\end{equation}

\begin{equation}
S_{\rm{ff}}(\nu) = 0.71 \nu^{-0.1} \frac{\rm{SFR}}{M_\odot \rm{yr}^{-1}} {D_{\rm{L}}}^{-2} \rm{Jy}
\label{eq:sff}
\end{equation}

\begin{equation}
S_{\rm{d}}(\nu) = 1.3\times10^{-6} \frac{\rm{SFR} (1+z) \nu^3}{{M_\odot \rm{yr}^{-1} D_{\rm{L}}^2}} e^{(0.048 \nu/T_{\rm{d}} -1)} \left[1-e^{-(\nu/2000)^{\beta}}\right] \rm{Jy}
\label{eq:sd}
\end{equation}

\begin{equation}
S_{\rm tot}(\nu)=S_{\rm ff}(\nu) + S_{\rm nth}(\nu) + S_{\rm d}(\nu) \rm{Jy}
\label{eq:tot}
\end{equation}

\noindent Our free parameters are therefore the non-thermal fraction ($f_{\rm nth}$), synchrotron spectral slope ($\alpha$) and star formation rate ($SFR$) with the dust emissivity index ($\beta$) and dust temperature ($T_{\rm d}$) fixed as previously mentioned.

Models are fit to the data using the Bayesian inference code MultiNest (\citealt{Feroz:08}) in combination with the Python wrapper PyMultinest \citep{Buchner:14} to solve for the free parameters. We assume a log-uniform prior for SFR (10-9000 M$_{\odot}$yr$^{-1}$) and uniform priors for log$_{10}f_{\rm nth}$ (0.5-1.5) and $\alpha$ (0.1-1.6) in order to best explore the full range of reasonable values for these parameters. PyMultiNest explores the parameter space using ellipsoidal nested sampling, returning a posterior probability distribution for each parameter and therefore full covariances for all parameters.

\begin{figure*}
\begin{center}
\begin{tabular}{cc}
\includegraphics[scale=0.5]{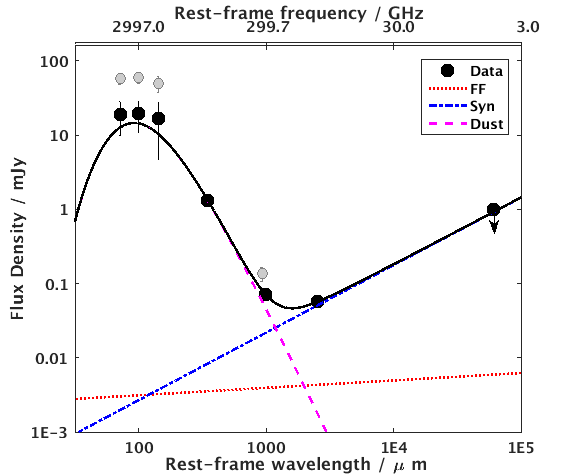} & \includegraphics[scale=0.5]{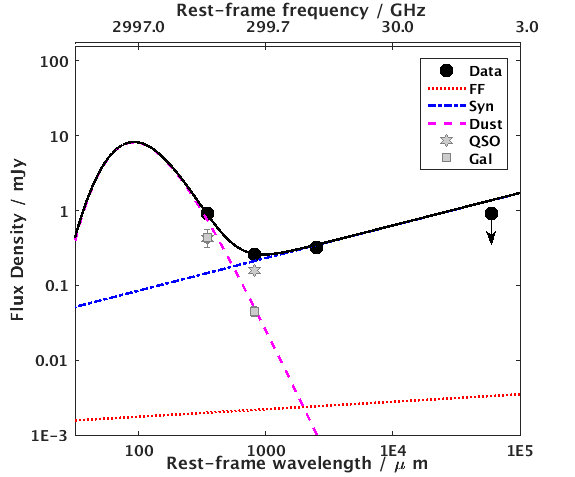}  \\
\end{tabular}
\caption{Best-fit SEDs to the far infra-red and radio continuum for ULASJ1234 (left) and ULASJ2315 (right). Black points represent the observed photometry used for the fitting. Error-bars for the ALMA and JVLA continuum data points are generally smaller than the symbol size on the plot. The SED is modelled as the sum of three components - a T$_{\rm{d}}=47$K, $\beta$=1.6 greybody to model the dust emission (dashed line), a power-law with spectral index, $\alpha$ to model synchrotron emission (dot-dashed line) and a free-free emission component with a power-law index of $\alpha_{\rm{ff}}=-0.1$ (dotted line). The free parameters in the fit are the SFR, the synchrotron spectral index, $\alpha$ and the scaling factor for the non-thermal synchrotron emission. We also show the upper limits on the 1.4GHz flux densities from FIRST for both quasars. In the case of ULASJ1234 we also include \textit{Herschel}-SPIRE photometry at 250, 350 and 500$\mu$m to constrain the peak of the SED. The total \textit{Herschel} emission shown by the grey points, is a blend of the quasar and the two other companion galaxies and for the fitting we assume that the fluxes are divided equally between the three systems (black points). We also show the total ALMA Band 3 continuum flux from all 3 galaxies in grey. In the case of ULASJ2315, the fits have been conducted by considering the integrated fluxes of both components of the major merger, as we do not have the angular resolution at all the observed frequencies to be able to disentangle emission from the two components of the merger. At the highest frequencies, where we do have enough angular resolution to be able to measure the flux densities for each component of the merger, we also show the individual flux densities for the QSO and the GAL component, although these have not been used in the fitting.}
\label{fig:seds}
\end{center}
\end{figure*}

The SED-fitting is carried out for ULASJ1234 and ULASJ2315 only where we have at least 3 continuum detections as well as upper limits from FIRST. As the \textit{Herschel} photometry at $\lambda_{\rm{obs}}$
=250, 350 and 500$\mu$m constrains the peak
of the dust SED, we also include this photometry in our SED-fitting for ULASJ1234 assuming equal contributions
from the three galaxies to the \textit{Herschel} fluxes. We have checked that dividing the 
\textit{Herschel} flux according to the ratio of ALMA fluxes in the three galaxies does not affect
the results significantly. The best-fit SEDs together with the individual components of the SED
for our two quasars, are presented in Fig. \ref{fig:seds}. The free-free contribution to the $\sim$33GHz
data points is negligible. Using Eq. \ref{eq:sff}, we estimate that SFRs of $\gtrsim$10$^4$M$_\odot$yr$^{-1}$
would be required for the free-free contribution to dominate at these frequencies. In the case of 
ULASJ2315, we caution that we have used the integrated flux
densities summed over both components of the merger as we lack the angular resolution necessary to be able to resolve the individual components of the
merger at all frequencies. However, in
Fig. \ref{fig:seds} we show for reference the derived flux densities
of both merger components at high frequencies where we are able to
separate out the contribution from the QSO and GAL.

In B17, we presented star formation rates and dust masses based on the
single continuum data point at $\nu_{\rm{obs}}\sim90-100$GHz. The new
continuum observations presented in this paper allow us to quantify
the non-thermal contribution to these frequencies and therefore revise
the star formation rates and dust masses. From our best-fit SEDs, we
find log$_{10}$(f$_{\rm{nth}}$)=1.2$\pm$0.3 and 1.3$\pm$0.1,
and SFRs of 1160$\pm$120 M$_\odot$ yr$^{-1}$ and
680$\pm$100 M$_\odot$yr$^{-1}$ for ULASJ1234 and ULASJ2315
respectively.  The values are consistent with the high SFRs inferred
for other FIR-luminous quasar host galaxies at high redshifts
\citep{Omont:01, Priddey:03}. Dust masses are calculated from the
dust SED components of these fits assuming the emission in the
Rayleigh-Jeans tail of the SED is optically thin:

\begin{equation}
M_{\rm{dust}}=\frac{D_{\rm{L}}^2}{(1+z)} \times \frac{S_{\nu,\rm{obs}}}{\kappa_{\nu,\rm{rest}} B(\nu_{\rm{rest}}, T_{\rm{d}})}
\label{eq:mdust}
\end{equation}

\noindent where $\kappa_{\nu,\rm{rest}}$ is the mass absorption
coefficient of the dust and is given by $\kappa_{0}(\nu/\nu_0)^\beta$,
$D_{\rm{L}}$ is the luminosity distance, $B(\nu_{\rm{rest}}, T_{\rm{d}})$ is the
Planck function and $\kappa_0$ is assumed to be 0.045 m$^2$ kg$^{-1}$
at $\nu_0$=250GHz. The new dust masses are
log$_{10}$(M$_{\rm{dust}}$)=8.32 and 8.09 for ULASJ1234 and ULASJ2315
respectively. In the case of ULASJ2315 our new observations show that non-thermal
emission dominates the $\nu_{\rm{obs}}\sim$100 GHz continuum used to
estimate the SFR and dust mass in B17. The new SFR and dust mass have therefore
decreased by an order of
magnitude relative to the results in B17. 

We have also constrained the synchrotron power-law index in both
quasars: $\alpha=0.9\pm0.2$ for ULASJ1234 and $\alpha=0.4\pm0.1$ in
ULASJ2315. The flat synchrotron power-law slope in ULASJ2315 almost
certainly arises because the quasar is dominating the
$\nu_{\rm{obs}}\sim$33 GHz JVLA continuum emission, whereas both the
quasar and the merging companion galaxy begin contributing to the
emission at higher frequencies (see Fig. \ref{fig:seds}). Higher
spatial resolution observations of the ULASJ2315 system at all
frequencies, would help constrain the exact form of the SED in the
individual components of the merger. Particularly for ULASJ2315, 
the limited amount of photometry available limits the constraints we can get
on the dust SED. We have therefore also fit the SEDs of both quasars after
fixing the synchrotron power-law index to values of $\alpha$=0.75 and $\alpha$=0.1, 
which encompass the full range expected in high-redshift quasars (e.g. \citealt{Momjian:14}). 
For ULASJ1234, a flat radio spectrum is ruled out by the data and the constraints on the 
SFR and the non-thermal fraction for $\alpha$=0.75 agree within the errors with the varying-
$\alpha$ fits. For ULASJ2315, the steep $\alpha$=0.75 SED is highly inconsistent with the 
FIRST upper limit and assuming $\alpha$=0.1 instead would reduce the SFR by a factor
of $\sim$1.5 to 430$\pm$90 M$_\odot$yr$^{-1}$. However, we again caution against 
over-interpretation of these numbers given the lack of constraints at the peak of the dust SED
and also the fact that we know this source is composed of two separate galaxies, which are not
resolved in our low-frequency data.

For ULASJ0123, where continuum emission at $\nu_{\rm{obs}}\sim$33 GHz
is detected with the JVLA, we do not have enough photometric data
points to be able to carry out the full SED-fitting described above.  Based on the
SEDs of ULASJ1234 and ULASJ2315 we assume that free-free emission
makes a negligible contribution to the SED, while synchrotron emission
dominates the JVLA continuum emission with a power-law index of
$\alpha=0.7$. Adopting such assumptions, we can estimate the
non-thermal emission contributing to the $\nu_{\rm{obs}}\sim$100 GHz
continuum used to infer star formation rates and dust masses in
B17. The result is $\sim$13 per cent, giving a revised SFR for the
quasar host galaxy of 2000$\pm$400M$\odot$yr$^{-1}$ (assuming
T$_{\rm{d}}$=47K and $\beta$=1.6) and a dust mass of
log$_{10}$(M$_{\rm{dust}}$/M$_\odot$)=8.63$\pm^{0.07}_{0.09}$. These
results are consistent, within the error-bars, with those derived in
B17. The results change by $<$10\% if a flatter synchrotron slope is instead assumed. 

Regardless of the exact constraints on the SFRs for these quasar hosts, 
which depend on the details of de-blending low-resolution observations of these multi-component sources as well as the assumptions made regarding the form of 
the SED, the inference of significant
amounts of star formation in all three quasar host galaxies is strongly evidenced by our data.

Finally, we have also detected dust continuum emission from the two
companion galaxies to ULASJ1234 - G1234S and G1234N (Table
\ref{tab:all_obs}). No continuum emission was detected at lower
frequencies with the JVLA for these sources, so it is reasonable to
assume that any contribution from non-thermal emission to the ALMA
data points is negligible. Estimates of the dust mass and star formation
rates in these companion galaxies depend heavily on the assumptions made
regarding the dust temperature and dust emissivity index in these galaxies. If 
we assume canonical values for starburst galaxies - T$_{\rm{d}}$=30K and
$\beta$=1.5 \citep{Casey:14}, the SFRs are $\sim$260 M$_\odot$yr$^{-1}$
and $\sim$190M$_\odot$yr$^{-1}$ for G1234S and G1234N respectively and 
the dust masses are log$_{10}$(M$_{\rm{dust}}$/M$_\odot$)$\sim$8.7-8.8. 
However, if a T$_{\rm{d}}$=47K and $\beta$=1.6 SED were assumed instead, 
as was done for the quasars, the SFRs would increase to $>$1000 M$_\odot$yr$^{-1}$
and the dust masses would be lower by $\sim$0.4 dex.

\subsection{Gas Excitation \& Physical Conditions in the ISM}

With multiple molecular lines detected in our reddened quasars, as
well as several of their companion galaxies, we can now try to better
understand the gas excitation and the properties of the interstellar
medium in these sources.

\subsubsection{The CO(3-2) to CO(1-0) Line Ratio}

\begin{figure}
\begin{center}
\includegraphics[scale=0.65]{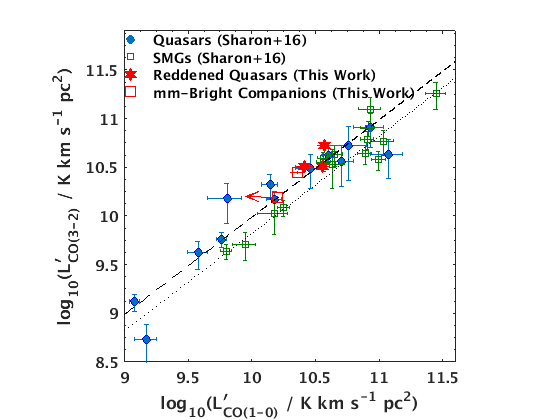}
\caption{The brightness temperature of the CO(3-2) line, L$^{\prime}_{\rm{CO(3-2)}}$ versus the brightness temperature of the CO(1-0) line, L$^{\prime}_{\rm{CO(1-0)}}$ for the sources studied in this paper, and compared to SMGs and quasars from \citet{Sharon:16}. In the latter case, all brightness temperatures have been corrected for magnification for lensed sources. The dashed line indicates a ratio between these brightness temperatures of r$_{31}$=0.97, which is considered to be an average value for quasar host galaxies, while the dotted line corresponds to r$_{31}$=0.66, which is an average value for SMGs \citep{Carilli:13}.}
\label{fig:r31}
\end{center}
\end{figure}

Combining the CO(1-0) fluxes for three of our quasars (Section
\ref{sec:CO10}) as well as the companion millimetre-bright galaxies to
ULASJ1234 (G1234N and G1234S) with the CO(3-2) detections in B17 and
Paper II, we can study the ratio of the brightness temperature between
these lines:
r$_{31}$=$L^{\prime}_{\rm{CO(3-2)}}/L^{\prime}_{\rm{CO(1-0)}}$.  While
historically there has been some evidence for a difference in r$_{31}$
between the quasar and SMG populations - \citet{Carilli:13} suggest
r$_{31}$=0.67 for SMGs and 0.99 for quasar host galaxies -
\citet{Sharon:16} presents evidence that quasars and SMGs have very
similar cold and warm gas reservoirs. In Fig. \ref{fig:r31} we plot $L^{\prime}_{\rm{CO(3-2)}}$ versus
$L^{\prime}_{\rm{CO(1-0)}}$ for our sources and compare them to both
quasars and SMGs from \citet{Sharon:16}, which are at a similar
redshift to our sample.

Our quasar host galaxies and their companion millimetre-bright
galaxies have CO(3-2) and CO(1-0) line emission that is fairly typical
of the high-redshift quasar and SMG sample and there is no evidence
for a statistical difference in excitation properties. We note,
however, that at least one of our sources - ULASJ2315 - is known to be
a close separation merger of two distinct galaxies. In Section
\ref{sec:CO_CI}, we have seen that the QSO and GAL components have
very different CO(7-6) and CI(2-1) line properties. Our observations
of CO(1-0) lack the angular resolution to separate emission from the
components.  However, in Section \ref{sec:CO10}, we have seen that the
peak of the CO(1-0) emission is spatially coincident with the
companion galaxy rather than the quasar host galaxy. If higher spatial
resolution observations of CO(1-0) indeed showed this to be the case,
and given that our high-resolution observations of CO(3-2) imply
comparable CO(3-2) luminosities in the QSO and GAL components (Paper II), 
this would imply that the QSO and SMG components of the merger have very
different values of r$_{31}$.

It is important to note that many of the published molecular gas
detections in SMGs and quasars e.g. with the VLA and Plateau de Bure
Interferometer (\citealt{Barvainis:98, Coppin:08, Carilli:02b}) lack
the spatial resolution to be able to resolve close separation mergers
such as ULASJ2315. Several of the hyper-luminous quasars and SMGs
studied in \citet{Sharon:16} for example, could plausibly be mergers
of quasars and SMGs and some dilution of any putative differences in
r$_{31}$ between quasars and SMGs is therefore expected when
considering emission averaged over multiple merging components with
different excitation properties.

\subsubsection{The CO(7-6) to CI(2-1) Line Ratio}

The CO(7-6) and CI(2-1) molecular emission lines have very different
critical densities and their line ratio can therefore be used to study
the relative amounts of high-density and low-density gas in the
ISM. Several recent studies have also suggested that in the high
cosmic-ray ionization rate environments that are thought to represent
high-redshift starburst and quasar-host galaxies, the CO molecule can
be destroyed and dissociate into atomic carbon \citep{Bisbas:15b},
leading to higher [CI]-to-CO line ratios. Comparing the relative line
intensities of CO(7-6) and CI(2-1), therefore provides us with an
effective diagnostic for such processes.

\begin{figure}
\begin{center}
\includegraphics[scale=0.65]{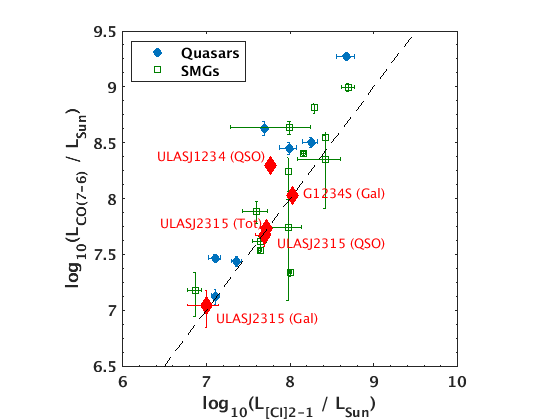}
\caption{The CO(7-6) versus CI(2-1) intrinsic (magnification corrected) luminosity for quasars and SMGs from \citet{Carilli:13} and \citet{Gullberg:16} compared to our reddened quasars and their companion galaxies. In the case of ULASJ2315, which is a close separation merger, we present both integrated line luminosities for the entire system as well as the individual line luminosities for each spatial component from Section \ref{sec:CO_CI}. The dashed line shows the 1:1 relation between the two luminosities.}
\label{fig:CO76_CI21}
\end{center}
\end{figure}

In order to put our galaxies in context, we have compiled a list of
quasars and SMGs at similar redshifts where both the CO(7-6) and
CI(2-1) lines have been detected \citep{Carilli:13,Gullberg:16}. We
derived intrinsic line luminosities for the sources in the literature
from the line intensities, using published redshifts and magnification
factors to correct for the effects of lensing. These line luminosities
are plotted in Fig. \ref{fig:CO76_CI21} alongside the measurements for
our reddened quasars and their companion galaxies. In the case of
ULASJ2315, we present both the integrated luminosities for the entire
system as well as the luminosities of the individual components. We find that while ULASJ1234
has a much higher CO(7-6) line luminosity compared to CI(2-1), the two line luminosities are 
fairly similar for both components of the ULASJ2315 major-merger system. Thus the amount
of high-density, high-excitation gas relative to the low-density gas appears to be
markedly different in the two reddened quasars. 
G1234S, the companion galaxy to ULASJ1234, also has a CO(7-6) line luminosity that
is comparable to the CI(2-1) line luminosity. 

\begin{figure*}
\begin{center}
\begin{tabular}{cc}
\includegraphics[scale=0.4]{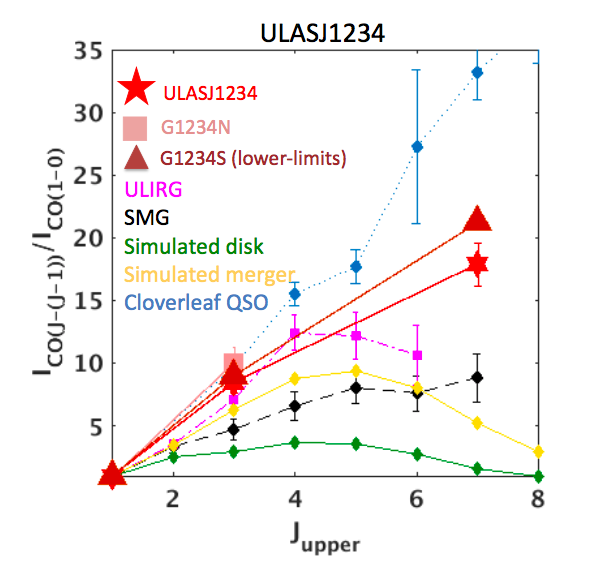} & \includegraphics[scale=0.4]{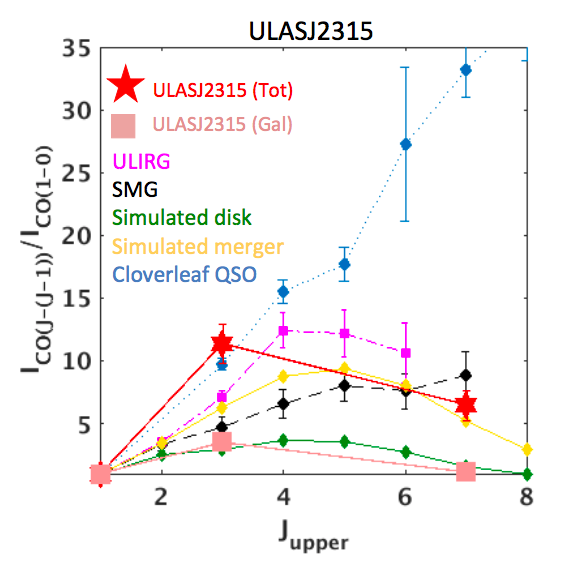}  \\
\end{tabular}
\caption{The CO Spectral Line Energy Distributions for all galaxies in the ULASJ1234 (left) and ULASJ2315 (right) systems. G1234N and G1234S are represented by the square and triangle symbols respectively in the left-hand panel, while the GAL component of ULASJ2315 is represented by the square symbols in the right-hand panel. As we did not detect CO(1-0) in G1234S, the triangles are used to represent lower limits on the line ratio for this galaxy. For ULASJ2315 we show integrated line ratios from both components of the merger as well as line ratios assuming all the CO(1-0) emission is coming from the northern companion galaxy. For comparison we also show the SLEDs for Ultraluminous Infrared Galaxies \citep{Papadopoulos:12}, Submillimetre Galaxies \citep{Bothwell:13}, the Cloverleaf quasar \citep{Carilli:13} and simulated SLEDs for both high-redshift disk galaxies and mergers \citep{Bournaud:15} that are designed to reproduce observational data for these kinds of systems.}
\label{fig:co_sleds}
\end{center}
\end{figure*}

\subsubsection{The CO SLED and Photo-Dissociation Region (PDR) Modelling}

\label{sec:pdr}

In Fig. \ref{fig:co_sleds} we present our constraints on the CO spectral
line energy distribution (SLED) for all galaxies in the ULASJ1234 and ULASJ2315
systems. The ULASJ1234 quasar host galaxy has a CO SLED peaking at $J>6$, 
which is inconsistent with even the most infrared luminous 
star-forming galaxies such as ULIRGs and SMGs \citep{Papadopoulos:12,Bothwell:13}
and could almost certainly be ascribed to heating from the luminous AGN. The gas 
excitation is not as extreme as seen in the Cloverleaf quasar for example, but lies 
somewhere between the quasar and SMG/ULIRG values. The
companion galaxy G1234S is similarly extreme in terms of its CO excitation properties. 
We did not detect CO(1-0) emission from this source but the CO(7-6) emission is 
considerably higher intensity compared to the CO(3-2) emission and based on the lines available, 
the CO SLED for this galaxy appears to be very similar to ULASJ1234. This strongly suggests
that X-ray or shock-heating could be important in this system even though we currently do not 
have any evidence for an AGN within it.

In ULASJ2315 on the other hand, the CO(1-0) emission does not spatially resolve the two
components of the major merger. Considering the integrated line properties from both galaxies suggests that the CO SLED peaks at lower-J compared to ULASJ1234 and the SLED is in fact consistent
with that seen in both ULIRGs and SMGs despite the presence of the luminous quasar. The
ULASJ2315 CO SLED, integrated across both components of the merger, is also in good agreement with simulations of galaxy mergers \citep{Bournaud:15}, which
were designed to match the observed excitation properties of distant galaxies. In Section \ref{sec:CO10} however we have seen that 
the CO(1-0) emission in ULASJ2315 is actually spatially co-incident with the companion galaxy rather than the quasar. If all the CO(1-0) emission is in fact coming from the companion galaxy, 
this would suggest that the galaxy, despite being part of the merger, has a CO SLED that is very 
similar to what is expected from quiescent disk galaxies (Fig. \ref{fig:co_sleds}; \citealt{Bournaud:15}). Thus even within this small, relatively
uniformly selected quasar sample, we are seeing evidence for a range in gas excitation properties.

We now use all available line
ratios for ULASJ1234, ULASJ2315 and G1234S, where we have detected the CO(3-2), CO(7-6) and CI(2-1) molecular line transitions, and compare them to the outputs of Photo Dissociation Region
(PDR) models. For ULASJ2315, we consider only the blended emission from the two galaxies in the merger as we do not have the necessary spatial resolution and 
S/N in all the lines to be able to fully disentangle the different spatial components of the line emission coming from the two distinct components of the merger. We
compare the observed line ratios to outputs from the 3D-PDR
code\footnote{https://uclchem.github.io/} \citep{Bisbas:12}. For
sources where we also have CO(1-0) line luminosities, we have checked
that the constraints do not change significantly when including the
CO(1-0) line in the fits. The 3D-PDR code is used to generate a grid
of models with uniform densities between $2 \le
\rm{log}(n/\rm{cm}^{-3}) \le 7$ and ultraviolet radiation field
strengths between $-0.2 \le \rm{log}(G/G_0) \le 5.8$ where $G_0$ is
the radiation field strength in Habing units. We assume a cosmic ray
(CR) ionization rate of 100$\times$ the Milky Way value -
i.e. $\zeta$=10$^{-15}$s$^{-1}$ - which is considered typical for
starburst galaxies, but we also assess how the results change for
models with $\zeta$=10$^{-17}$, 5$\times$10$^{-15}$ and 10$^{-14}$
s$^{-1}$. All outputs from the model are integrated line luminosities
down to a depth of A$_{\rm{V}}$=7 mag, which should be representative
of the average properties of star-forming clouds seen in our high
redshift galaxies \citep{Pelupessy:06}. Qualitatively, the results are
very similar for depths of A$_{\rm{V}}$=2 and 3 mag.

For G1234S, no constraints on the gas density and radiation field strength were
possible even after exploring the full parameter space of the PDR models. As stated 
earlier, the very high excitation gas seen in this galaxy could indicate that X-ray or mechanical heating 
dominate over far ultraviolet heating in this system. As this physics is currently not incorporated into our
PDR models we are unable to constrain the ISM properties of this galaxy. The constraints on the gas densities and radiation field strength for
different values of $\zeta$ for the two quasars are given in Table
\ref{tab:pdr_outputs}. In general we find high gas densities of
$\sim$10$^5$-10$^6$ cm$^{-3}$ in both sources. The available line
ratios do not constrain the radiation field strength as well as the
density. The exact constraints on the radiation field strength also
depend strongly on the CR-ionization rate; lower radiation field
strengths are generally preferred for higher CR-ionization rate
models as the effect of both is to raise the gas temperature in the ISM. 
For ULASJ1234 CR-ionization rates as high
as 10$^{-14}$s$^{-1}$ are allowed by the data. For ULASJ2315 however, such high
ionization rate models are excluded and the gas in the galaxy is
likely at lower temperatures.

\begin{table*}
\caption{Constraints on the gas density, $n$ and radiation field strength $G$ of the ISM in our reddened quasars ULASJ1234 and ULASJ2315, obtained by comparing the observed line ratios to PDR model outputs generated using the 3D-PDR code. Constraints are given for models with a range of cosmic ray ionization rates, $\zeta$, and by integrating down to a depth of A$_{\rm{V}}$=7 mag in the models. We also provide the predicted brightness temperature of the CI(1-0) line, L$^{\prime \rm{pred}}_{\rm{CI(1-0)}}$ from the models, which is used later in the paper.}
\label{tab:pdr_outputs}
\begin{center}
\begin{tabular}{lccccc}
Source & A$_{\rm{V}}$ / mag & $\zeta$ / s$^{-1}$ & $\rm{log}(n/\rm{cm}^{-3})$ & $\rm{log}(G/G_0)$ / & L$^{\prime \rm{pred}}_{\rm{CI(1-0)}} \times10^9$ K km s$^{-1}$ pc$^2$\\
\hline
ULASJ1234+0907 (QSO) & 7 & 10$^{-17}$ & 6.33$\pm^{0.08}_{0.13}$ & 3.25$\pm^{0.15}_{0.15}$ & 3.4$\pm$0.4 \\
& 7 & 10$^{-15}$ & 6.12$\pm^{0.18}_{0.12}$ & 4.58$\pm^{1.33}_{1.28}$ & 3.7$\pm$0.5 \\
& 7 & 5$\times$10$^{-15}$ & 5.99$\pm^{0.21}_{0.59}$ & 2.55$\pm^{1.33}_{1.28}$ & 4.6$\pm$1.3 \\
& 7 & 10$^{-14}$ & 5.86$\pm^{0.34}_{0.36}$ & 1.33$\pm^{1.17}_{1.33}$ & 5.6$\pm$1.4 \\
ULASJ2315+0143 (QSO+Gal) & 7 & 10$^{-17}$ & 5.70$\pm^{0.01}_{0.01}$ & 1.95$\pm^{0.15}_{0.15}$ & 4.4$\pm$0.6 \\
& 7 & 10$^{-15}$ & 5.20$\pm^{0.10}_{0.10}$ & 1.45$\pm^{0.25}_{0.25}$ & 6.0$\pm$1.0\\
\hline
\end{tabular}
\end{center}
\end{table*}

\begin{figure*}
\begin{center}
\includegraphics[scale=0.7]{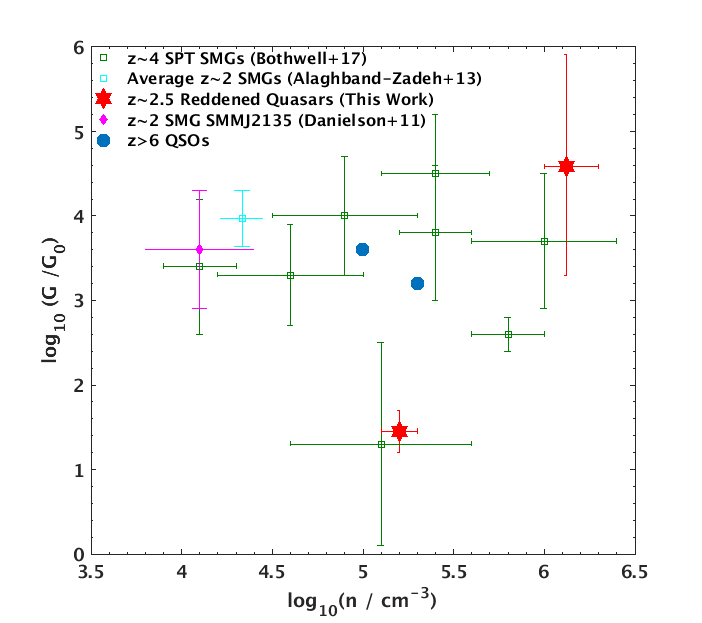}
\caption{The gas density versus radiation field strength inferred from PDR modelling of our two heavily reddened quasars, ULASJ1234 and ULASJ2315 assuming $\zeta$=10$^{-15}$s$^{-1}$ and A$_{\rm{V}}$=7 mags. These values are compared to both SMGs and quasars from the literature \citep{Alaghband:13, Danielson:11, Bothwell:17, Venemans:17}. In the case of ULASJ2315, we plot constraints from the integrated emission from the QSO and Gal components of the merger. The \citet{Bothwell:17} SMG sample forms the most direct comparison sample to ours, on account of using exactly the same PDR model and model set-up as assumed in this work.}
\label{fig:logn_logG}
\end{center}
\end{figure*}

In Fig. \ref{fig:logn_logG} we compare the inferred gas densities and
radiation field strengths in our two sources for a fiducial CR ionization
rate, $\zeta$=10$^{-15}$s$^{-1}$, to other high-redshift quasars and
SMGs from the literature. \citet{Bothwell:17} provides the most direct
comparison SMG sample to ours, employing the same PDR-code and model
set-up to calculate line ratios. In general our quasars appear to have physical conditions typical of both
SMGs and other quasars from the literature. Both quasar host galaxies have
high gas densities although they exhibit very different radiation field strengths. 
Based on our best-fit PDR models, we also derive an
excitation temperature for atomic carbon, which then allows us to
predict the CI(1-0) line brightness temperature. These values are
quoted in Table \ref{tab:pdr_outputs} and used later in the paper.

As hinted at earlier, a caveat of our current modelling is that the 3D-PDR code does not
include X-ray and mechanical heating, the effects of which might be expected to be significant 
in our merging quasar host galaxies. Certainly in the case of ULASJ1234 and G1234S, X-ray heating
could be responsible for the high values of the CO(7-6) intensity that we observe in
these sources. Observations of other molecular lines such as HCN and HCO+, when combined
with the high-J CO lines, could help discriminate between cosmic-ray heating and X-ray heating
\citep{Meijerink:11}. For ULASJ2315 on the other hand, the similarity of the CO SLED to that of non-AGN
populations such as SMGs and ULIRGs, suggests that the effects of X-ray heating are probably not as significant.

Mechanical heating due to shocks from supernova remnants as well as the effects of supersonic turbulence
could also contribute to the gas heating and explain the highly excited CO SLEDs seen in some of our sources 
\citep{Papadopoulos:12}. Once again, observations of other lines are necessary to be able to distinguish these
processes from the far ultra-violet heating that dominates in the PDR models.

\subsection{Mass of Atomic Carbon}

\label{sec:carbon}

The CI(2-1) detections presented in Section \ref{sec:CO_CI} now add to
a growing number of sources in the literature where atomic carbon has
been detected \citep{Weiss:05, Walter:11, Alaghband:13, Wang:16,
Bothwell:17, Venemans:17}. Due to the relative simplicity of the [CI]
quantum fine-structure levels compared to the widely studied
rotational transitions of the CO molecule, physical parameters
such as the excitation temperature and total mass of carbon can be
calculated with fewer uncertainties.

We now estimate the total mass of atomic carbon from the CI(2-1)
line luminosities \citep{Weiss:03}

\begin{equation}
\rm{M}_{\rm{CI}}=4.566 \times 10^{-4} Q(T_{\rm{ex}}) \frac{1}{5} e^{T_2/T_{\rm{ex}}}L'_{\rm{[CI]}^3P_2-^3P_1}
\label{eq:mci}
\end{equation}

\noindent where
$Q(T_{\rm{ex}})=1+3e^{-T_1/T_{\rm{ex}}}+5e^{-T_2/T_{\rm{ex}}}$ is the
[CI] partition function, $T_1$=23.6K and T$_2$=62.5K are the energies
above the ground state. We assume that the line emission from both CI
lines is optically thin, and that the lines are in local thermodynamic
equilibrium (LTE). We calculate the excitation temperature from the
[CI]$^3P_2-^3P_1$ to [CI]$^3P_1-^3P_0$ brightness temperature ratios
calculated from our best-fit PDR models with
$\zeta$=10$^{-15}$s$^{-1}$ and A$_{\rm{V}}$=7 mags for each of the
galaxies:

\begin{equation}
T_{\rm{ex}}=38.8/ln\left(\frac{2.11 L'_{\rm{[CI]}^3P_1-^3P_0}}{L'_{\rm{[CI]}^3P_2-^3P_1}}\right)
\label{eq:Tex}
\end{equation}

\noindent These excitation temperatures range from $\sim$30K in ULASJ2315 to $\sim$50 K in ULASJ1234 and are
consistent with the values observed in other high-redshift galaxies
where both the CI(2-1) and CI(1-0) lines have been detected
\citep{Weiss:03, Walter:11}. The total mass of carbon is relatively
insensitive to the exact value of the excitation temperature for
T$_{\rm{ex}}\gtrsim$20K. We find $\rm{M}_{\rm{CI}}$=
(4.9$\pm$0.6)$\times$10$^6$M$_\odot$ for ULASJ1234 and (8.0$\pm$1.3)$\times$10$^6$M$_\odot$
for the total ULASJ2315 system. For G1234S PDR model-fits aren't available to directly infer the CI excitation temperature. We therefore
assume the average
value for high-redshift galaxies from \citet{Walter:11}. Under these assumptions, we find 
an atomic carbon mass of (6.2$\pm$0.8)$\times$10$^6$M$_\odot$ for this galaxy. 

\subsection{Gas Masses from CO, [CI] and Dust: Implications for the gas-to-dust ratio, carbon abundance and $\alpha_{\rm{CO}}$}

We have measured the dust continuum emission as well as emission from
multiple molecular lines in our high-redshift quasars and their
companion galaxies, which has provided direct constraints on the gas
excitation in these systems. We can therefore compare gas masses
estimated using different tracers in order to test the validity of the
assumptions that are implicit in these various gas-mass derivations.

\subsubsection{Reconciling CO(3-2) and CO(1-0) based Gas Masses}

In B17, we derived gas masses from the brightness temperature of the
CO(3-2) line, L$^{\prime}_{\rm{CO(3-2)}}$ assuming an r$_{31}$ value
of 0.8 for the quasar host galaxies and r$_{31}$=0.67 for the
companions \citep{Carilli:13}. With both CO(1-0) and CO(3-2)
measurements now available, we can directly measure r$_{31}$ and
derive gas masses from the CO(1-0) emission. The constraints on
r$_{31}$ are given in Table \ref{tab:gas_mass}. For all 3 quasars and both companion galaxies
in the ULASJ1234 system, we find $r_{31}\gtrsim1$, although we
caution that for ULASJ2315, the interpretation of the brightness
temperature ratio is almost certainly complicated by the fact
that this source is part of a major merger with likely very different
excitation conditions in the merging galaxies. Nevertheless, the gas
masses inferred from the CO(1-0) emission would be lower than those
inferred in B17 using CO(3-2). Values of r$_{31}>1$ do not usually
occur in high-redshift starburst galaxies, where the ISM is thought to
be dominated by molecular gas and the CO emission is optically
thick. Given the relatively modest S/N of the CO(1-0) detections, it
is likely that we are missing extended, low surface-brightness
emission below our detection threshold. Alternatively, the
superthermal values of r$_{31}$, particularly in the case of ULASJ0123
where we find r$_{31}=1.4$, could also result from X-ray heating or
shock heating of the interstellar gas, which can penetrate much deeper
into molecular clouds and therefore maintain a low optical depth in CO
emission \citep{Meijerink:06, Ivison:12}. Such mechanisms could
plausibly be at work in our high-luminosity quasars.

In Table \ref{tab:gas_mass}, we present gas masses
derived from the CO(1-0) as well as the CO(3-2) brightness temperature
for all sources. In cases where
L$^{\prime}_{\rm{CO(3-2)}}>$L$^{\prime}_{\rm{CO(1-0)}}$ or where we
currently have no constraints on L$^{\prime}_{\rm{CO(1-0)}}$, we
assume r$_{31}$=1.0.

\subsubsection{Dust-based Gas Masses}

In Section \ref{sec:cont} we have quantified the contributions from
non-thermal emission to the observed continuum fluxes and therefore
better constrained the dust SED and the dust masses in these quasars
compared to those presented in B17. With some knowledge of the
gas-to-dust ratios in our high-redshift galaxies, the dust masses can
be used as a proxy for the gas masses \citep{Scoville:14}. In B17 we assumed that a gas-to-dust ratio of 91, as measured in
nearby galaxies \citep{Sandstrom:13}, was appropriate for our
high-redshift galaxies. In this work, we allow for a range of gas-to-dust ratios given that this quantity is known to depend on metallicity
\citep{Draine:07, Remy-Ruyer:14}. We take the gas-phase metallicities of high-redshift star-forming galaxies from 
\citet{Santini:14} to be representative of our high-redshift sample. The range in metallicities - 12+log(O/H)=8.58-9.07 - translates to 
gas-to-dust ratios of 42-130 based on \citet{Draine:07}, with a mean value of 62 corresponding to a mean gas-phase metallicity of 8.9 for the star-forming galaxies.
We then derive gas masses from the dust
masses for all our quasars and companion galaxies assuming these gas-to-dust ratios. We have
deliberately not constrained the dust masses or the dust-based gas
masses for the two companion galaxies to ULASJ1234,G1234N and G1234S, as we only have a limited amount of photometry and no 
knowledge of the form of the dust SED in these sources. Hence any dust-based gas mass estimates would be highly uncertain. The
dust masses as well as the dust-based gas masses are summarised in
Table \ref{tab:gas_mass}. 

There is a large uncertainty in the dust-based gas masses when allowing
for a range in metallicities in the sample. Nevertheless,  
the CO-based gas masses are higher than those predicted from the dust
for both ULASJ1234 and ULASJ2315. This could imply values of
$\alpha_{\rm{CO}}<0.8$ M$_\odot$(K kms$^{-1}$pc$^2$)$^{-1}$, or,
alternatively, gas-to-dust ratios of $\sim$150-200, which would imply lower
metallicities than observed in typical star-forming galaxies at these redshifts.

\subsubsection{Gas Masses from Atomic Carbon}

It has been argued by several authors that the CI emission can be a
more effective tracer of the total gas mass in high-redshift galaxies
compared to CO, particularly at low metallicities and for high cosmic
ray ionization rates where the CO molecule is known to become an
increasingly poorer tracer of total gas mass \citep{Glover:16}. In
Section \ref{sec:pdr} we used the results of our PDR modelling
together with the observed brightness temperature of the CI(2-1) line
in our sources (Table \ref{tab:all_obs}), to predict the CI(1-0)
brightness temperature, L$^{\prime}_{\rm{[CI]}^3P_1-^3P_0}$. We now
use these brightness temperatures to calculate a CI-based gas mass
using the relation:

\begin{equation}
\rm{M(H_2)^{CI}}=3.97\times10^{-5} {\nu_{\rm{rest}}}^2 \frac{10^{-5}}{X_{\rm{CI}}} \frac{10^{-7}}{A_{10}} \frac{1}{Q_{10}} L'_{\rm{[CI]}^3P_1-^3P_0}
\label{eq:mgas_ci}
\end{equation}

\noindent where X$_{\rm{CI}}$=CI/H$_2$ represents the atomic carbon
abundance, A$_{10}$=7.93$\times$10$^{-8}$s$^{-1}$ is the Einstein A
coefficient and Q$_{10}$ is the excitation factor, which depends on
the temperature, density and radiation field of the ISM. We assume a
value of X$_{\rm{CI}}$=3$\times10^{-5}$, based on a template of M82
used previously for high-redshift SMGs and quasars \citep{Weiss:03,
Weiss:05}. Q$_{10}$ is assumed to be 0.6 \citep{Bothwell:17}. The
resulting gas masses are once again presented in Table
\ref{tab:gas_mass}. In ULASJ1234 the CI-based gas mass is
consistent with the CO-based gas mass and the inferred values of
$\alpha_{\rm{CO}}$ and the atomic carbon abundance are therefore
consistent with the canonical values assumed in our gas mass
derivations. However, for ULASJ2315, we find a CI-based gas
mass that is a factor of $\sim$2 higher than that inferred from CO. Given
the uncertainties on each measurement, this difference is significant at the 
$\sim$2$\sigma$ level. In G1234S however, the CI-based gas mass estimate is almost 6$\times$ higher than
that inferred from CO. The CI and CO-based gas masses can be reconciled if
$\alpha_{\rm{CO}}>>$0.8 M$_\odot$(K kms$^{-1}$pc$^2$)$^{-1}$, or,
alternatively, if the carbon abundance in G1234S is $\sim$6$\times$ higher
than the assumed value of 3$\times$10$^{-5}$.

\begin{table*}
\begin{center}
\caption{Comparison of gas masses estimated using different tracers for the heavily reddened quasars and companion galaxies in our sample and the resulting constraints on r$_{31}$, $\alpha_{\rm{CO}}$, the gas-to-dust ratio and the atomic carbon abundances. Estimates of the star formation rates for each galaxy, calculated from the continuum photometry, are also quoted. For both the dust masses and star formation rates the uncertainties are dominated by our incomplete knowledge of the dust SED in these systems given the limited photometric data. We have therefore quoted these numbers without formal uncertainties.}
\label{tab:gas_mass}
\begin{tabular}{cccccc}
& ULASJ0123 & ULASJ1234 & G1234N & G1234S & ULASJ2315 (Tot) \\
\hline
SFR / M$_\odot$ yr$^{-1 \circ}$ & 2000 & 1160 & 190 & 260 & 680 \\ 
\hline
CO(1-0) M$_{\rm{gas}}^{\ast}$ / 10$^{10}$ M$_\odot$ & 3.0$\pm$0.4 & 2.9$\pm$0.3 & 1.8$\pm$0.2 & $<$1.3 & 2.1$\pm$0.3  \\
r$_{31}$ & 1.4$\pm$0.2 & 0.9$\pm$0.1 & 1.2$\pm$0.1 & $>$0.98 & 1.2$\pm$0.2 \\
CO(3-2) M$_{\rm{gas}}^{\star}$ / 10$^{10}$ M$_\odot$ & 4.2$\pm$0.2 & 2.9$\pm$0.3 & 2.2$\pm$0.2 & 1.25$\pm$0.01 & 2.6$\pm$0.2 \\
\hline
log$_{10}$(M$_{\rm{dust}}$/M$_\odot$)$^{\circ}$ & 8.63 & 8.32 & 8.66 & 8.75 & 8.09 \\
Dust M$_{\rm{gas}}^{\dagger}$ / 10$^{10}$ M$_\odot$ & 1.8-5.5 & 0.9-2.7 & -- & -- & 0.5-1.6 \\
$\alpha_{\rm{CO}}^{\rm{dust}}$ / M$_\odot$ (K kms$^{-1}$ pc$^2$)$^{-1}$ & 0.5-1.5 & 0.2-0.8 & -- & -- & 0.2-0.6 \\
Gas-to-dust ratio & 70$-$100 & 140 & -- & -- & 170$-$210 \\ 
\hline
CI M$_{\rm{gas}}^{\ddagger}$ / 10$^{10}$ M$_\odot$ & -- & 2.5$\pm$0.3 & -- & 7.3$\pm$0.4 & 4.1$\pm$0.7 \\
$\alpha_{\rm{CO}}^{\rm{CI}}$ / M$_\odot$ (K kms$^{-1}$ pc$^2$)$^{-1}$  & -- & 0.7 & -- & $>$4.6 & 1.6 \\
$\rm{X}_{\rm{CI}}$ / $\times$10$^{-5}$ & -- & 2.8 & -- & 18.3 & 5.1$-$6.1 \\ 
\hline
\end{tabular}
\end{center}
\flushleft
\small{$^{\circ}$Assuming T$_{\rm{d}}$=47K and $\beta$=1.6 for ULASJ0123, ULASJ1234 and ULASJ2315 and T$_{\rm{d}}$=30K and $\beta$=1.5 for G1234N and G1234S.} \\
\small{$^{\ast}$Assuming $\alpha_{\rm{CO}}$=0.8 M$_\odot$ (K km s$^{-1}$ pc$^2$)$^{-1}$} \\
\small{$^{\star}$Assuming r$_{31}$=1.0 when L$^{\prime}_{\rm{CO(3-2)}}>$L$^{\prime}_{\rm{CO(1-0)}}$ or there is no constraint on L$^{\prime}_{\rm{CO(1-0)}}$, and derived value of r$_{31}$ otherwise. Also assumes $\alpha_{\rm{CO}}$=0.8 M$_\odot$ (K km s$^{-1}$ pc$^2$)$^{-1}$} \\
\small{$^{\dagger}$Assuming a metallicity-dependent gas-to-dust ratio of 42-130 and based on the range in gas-phase metallicities seen in high-redshift star-forming galaxies.} \\
\small{$^{\ddagger}$Assuming an atomic carbon abundance of 3$\times$10$^{-5}$} 
\end{table*}

\section{Conclusions}

\label{sec:conclusions}

We have presented JVLA and ALMA observations of the CO(1-0), CO(7-6)
and CI(2-1) molecular lines as well as the far-infrared to radio
continuum in three heavily reddened quasars and their associated
companion galaxies at $z\sim2.5$. Combining these observations with
our previous observations of CO(3-2) emission allows us to study in
detail the properties of the interstellar medium and the gas
excitation conditions. We find that our quasar host galaxies as well
as their companion galaxies have ISM properties consistent with the
presence of high-density, high-temperature gas that is already highly
enriched in elements like carbon. We detect CI(2-1) emission tracing
low-density, diffuse gas in several of our galaxies and compare this to
the CO(7-6) line luminosities, which trace dense, more highly excited 
gas. In general the CO(7-6) and CI(2-1) line luminosities are comparable. 
The CO spectral line energy distributions even in this small sample span 
the full range in gas excitation conditions from luminous starburst galaxies
to the brightest quasars. 

One of our quasar host galaxies (ULASJ2315) is spatially resolved into
a close-separation ($<$2$\arcsec$) major merger and we present some evidence that the
gas excitation conditions vary significantly between the two galaxies.
Many previous studies have used low angular resolution observations of
similar systems to infer the average line ratios and excitation
conditions in high-redshift galaxies and we therefore caution that
such results could be biased by potentially averaging over multiple
merging components with different properties.

We derive gas masses for our galaxies in three different ways - (i)
using CO emission (ii) using dust emission and (iii) using CI
emission. We demonstrate that for standard assumptions for the
CO-to-H$_2$ conversion factor, gas-to-dust ratio and atomic carbon
abundance, these estimates sometimes do not agree. The differences in
the gas-mass estimates could indicate a range of $\alpha_{\rm{CO}}$
values, or, alternatively, high gas-to-dust ratios and enhanced
atomic carbon abundances.

In general, our investigations highlight the diversity in the dust and
gas distributions and ISM properties of high-redshift quasar host
galaxies and their companion starburst galaxies at the main epoch of
galaxy formation. With ALMA now opening up the window on
high angular resolution observations of dust and gas at
high-redshifts, we can begin to move from inferring globally averaged
properties to understanding spatial variations in the ISM properties
in these complex, merging systems.

\section*{Acknowledgements}

We thank the anonymous referee for their careful reading and useful comments on the paper. We thank Matt Bothwell and Stefano Carniani for useful discussions. MB acknowledges funding from the UK Science and Technology Facilities Council (STFC) via an Ernest Rutherford Fellowship and from the Royal Society via a University Research Fellowship. GCJ acknowledges support from NRAO through the Grote Reber Doctoral Fellowship Program. PCH acknowledges funding from STFC via the Institute of Astronomy, Cambridge, Consolidated Grant.

This paper makes use of the following ALMA data: ADS/JAO.ALMA\#2016.1.01308.S. ALMA is a partnership of ESO (representing its member states), NSF (USA) and NINS (Japan), together with NRC (Canada), NSC and ASIAA (Taiwan), and KASI (Republic of Korea), in cooperation with the Republic of Chile. The Joint ALMA Observatory is operated by ESO, AUI/NRAO and NAOJ.

The National Radio Astronomy Observatory is a facility of the National Science Foundation operated under cooperative agreement by Associated Universities, Inc.




\bibliography{}

\begin{thebibliography}{}

\bibitem[\protect\citeauthoryear{{Alaghband-Zadeh}, {Chapman}, {Swinbank},
  {Smail}, {Danielson}, {Decarli}, {Ivison}, {Meijerink}, {Weiss} \& {van der
  Werf}}{{Alaghband-Zadeh} et~al.}{2013}]{Alaghband:13}
{Alaghband-Zadeh} S.,  {Chapman} S.~C.,  {Swinbank} A.~M.,  {Smail} I.,
  {Danielson} A.~L.~R.,  {Decarli} R.,  {Ivison} R.~J.,  {Meijerink} R.,
  {Weiss} A.,    {van der Werf} P.~P.,  2013, \mnras, 435, 1493

\bibitem[\protect\citeauthoryear{{Banerji}, {Carilli}, {Jones}, {Wagg},
  {McMahon}, {Hewett}, {Alaghband-Zadeh} \& {Feruglio}}{{Banerji}
  et~al.}{2017}]{Banerji:17}
{Banerji} M.,  {Carilli} C.~L.,  {Jones} G.,  {Wagg} J.,  {McMahon} R.~G.,
  {Hewett} P.~C.,  {Alaghband-Zadeh} S.,    {Feruglio} C.,  2017, \mnras, 465,
  4390

\bibitem[\protect\citeauthoryear{{Banerji} et~al.,}{{Banerji}
  et~al.}{2015}]{Banerji:15}
{Banerji} M.,  et~al., 2015, \mnras, 447, 3368

\bibitem[\protect\citeauthoryear{{Banerji}, {Fabian} \& {McMahon}}{{Banerji}
  et~al.}{2014}]{Banerji:14}
{Banerji} M.,  {Fabian} A.~C.,    {McMahon} R.~G.,  2014, \mnras, 439, L51

\bibitem[\protect\citeauthoryear{{Banerji}, {McMahon}, {Hewett},
  {Alaghband-Zadeh}, {Gonzalez-Solares}, {Venemans} \& {Hawthorn}}{{Banerji}
  et~al.}{2012}]{Banerji:12}
{Banerji} M.,  {McMahon} R.~G.,  {Hewett} P.~C.,  {Alaghband-Zadeh} S.,
  {Gonzalez-Solares} E.,  {Venemans} B.~P.,    {Hawthorn} M.~J.,  2012, \mnras,
  427, 2275

\bibitem[\protect\citeauthoryear{{Banerji}, {McMahon}, {Hewett},
  {Gonzalez-Solares} \& {Koposov}}{{Banerji} et~al.}{2013}]{Banerji:13}
{Banerji} M.,  {McMahon} R.~G.,  {Hewett} P.~C.,  {Gonzalez-Solares} E.,
  {Koposov} S.~E.,  2013, \mnras, 429, L55

\bibitem[\protect\citeauthoryear{{Barvainis}, {Alloin}, {Guilloteau} \&
  {Antonucci}}{{Barvainis} et~al.}{1998}]{Barvainis:98}
{Barvainis} R.,  {Alloin} D.,  {Guilloteau} S.,    {Antonucci} R.,  1998,
  \apjl, 492, L13

\bibitem[\protect\citeauthoryear{{Beelen}, {Cox}, {Benford}, {Dowell},
  {Kov{\'a}cs}, {Bertoldi}, {Omont} \& {Carilli}}{{Beelen}
  et~al.}{2006}]{Beelen:06}
{Beelen} A.,  {Cox} P.,  {Benford} D.~J.,  {Dowell} C.~D.,  {Kov{\'a}cs} A.,
  {Bertoldi} F.,  {Omont} A.,    {Carilli} C.~L.,  2006, \apj, 642, 694

\bibitem[\protect\citeauthoryear{{Bisbas}, {Bell}, {Viti}, {Yates} \&
  {Barlow}}{{Bisbas} et~al.}{2012}]{Bisbas:12}
{Bisbas} T.~G.,  {Bell} T.~A.,  {Viti} S.,  {Yates} J.,    {Barlow} M.~J.,
  2012, \mnras, 427, 2100

\bibitem[\protect\citeauthoryear{{Bisbas}, {Papadopoulos} \& {Viti}}{{Bisbas}
  et~al.}{2015}]{Bisbas:15b}
{Bisbas} T.~G.,  {Papadopoulos} P.~P.,    {Viti} S.,  2015, \apj, 803, 37

\bibitem[\protect\citeauthoryear{{Bothwell} et~al.,}{{Bothwell}
  et~al.}{2017}]{Bothwell:17}
{Bothwell} M.~S.,  et~al., 2017, \mnras, 466, 2825

\bibitem[\protect\citeauthoryear{{Bothwell}, {Smail}, {Chapman}, {Genzel},
  {Ivison}, {Tacconi}, {Alaghband-Zadeh}, {Bertoldi}, {Blain}, {Casey}, {Cox},
  {Greve}, {Lutz}, {Neri}, {Omont} \& {Swinbank}}{{Bothwell}
  et~al.}{2013}]{Bothwell:13}
{Bothwell} M.~S.,  {Smail} I.,  {Chapman} S.~C.,  {Genzel} R.,  {Ivison} R.~J.,
   {Tacconi} L.~J.,  {Alaghband-Zadeh} S.,  {Bertoldi} F.,  {Blain} A.~W.,
  {Casey} C.~M.,  {Cox} P.,  {Greve} T.~R.,  {Lutz} D.,  {Neri} R.,  {Omont}
  A.,    {Swinbank} A.~M.,  2013, \mnras, 429, 3047

\bibitem[\protect\citeauthoryear{{Bournaud}, {Daddi}, {Wei{\ss}}, {Renaud},
  {Mastropietro} \& {Teyssier}}{{Bournaud} et~al.}{2015}]{Bournaud:15}
{Bournaud} F.,  {Daddi} E.,  {Wei{\ss}} A.,  {Renaud} F.,  {Mastropietro} C.,
   {Teyssier} R.,  2015, \aap, 575, A56

\bibitem[\protect\citeauthoryear{{Brusa} et~al.,}{{Brusa}
  et~al.}{2015}]{Brusa:15}
{Brusa} M.,  et~al., 2015, \aap, 578, A11

\bibitem[\protect\citeauthoryear{{Buchner} et~al.,}{{Buchner}
  et~al.}{2014}]{Buchner:14}
{Buchner} J.,  et~al., 2014, \aap, 564, A125

\bibitem[\protect\citeauthoryear{{Carilli}, {Cox}, {Bertoldi}, {Menten},
  {Omont}, {Djorgovski}, {Petric}, {Beelen}, {Isaak} \& {McMahon}}{{Carilli}
  et~al.}{2002}]{Carilli:02b}
{Carilli} C.~L.,  {Cox} P.,  {Bertoldi} F.,  {Menten} K.~M.,  {Omont} A.,
  {Djorgovski} S.~G.,  {Petric} A.,  {Beelen} A.,  {Isaak} K.~G.,    {McMahon}
  R.~G.,  2002, \apj, 575, 145

\bibitem[\protect\citeauthoryear{{Carilli}, {Hodge}, {Walter}, {Riechers},
  {Daddi}, {Dannerbauer} \& {Morrison}}{{Carilli} et~al.}{2011}]{Carilli:11}
{Carilli} C.~L.,  {Hodge} J.,  {Walter} F.,  {Riechers} D.,  {Daddi} E.,
  {Dannerbauer} H.,    {Morrison} G.~E.,  2011, \apjl, 739, L33

\bibitem[\protect\citeauthoryear{{Carilli}, {Kohno}, {Kawabe}, {Ohta},
  {Henkel}, {Menten}, {Yun}, {Petric} \& {Tutui}}{{Carilli}
  et~al.}{2002}]{Carilli:02}
{Carilli} C.~L.,  {Kohno} K.,  {Kawabe} R.,  {Ohta} K.,  {Henkel} C.,  {Menten}
  K.~M.,  {Yun} M.~S.,  {Petric} A.,    {Tutui} Y.,  2002, \aj, 123, 1838

\bibitem[\protect\citeauthoryear{{Carilli} \& {Walter}}{{Carilli} \&
  {Walter}}{2013}]{Carilli:13}
{Carilli} C.~L.,  {Walter} F.,  2013, \araa, 51, 105

\bibitem[\protect\citeauthoryear{{Casey}, {Narayanan} \& {Cooray}}{{Casey}
  et~al.}{2014}]{Casey:14}
{Casey} C.~M.,  {Narayanan} D.,    {Cooray} A.,  2014, \physrep, 541, 45

\bibitem[\protect\citeauthoryear{{Condon}}{{Condon}}{1992}]{Condon:92}
{Condon} J.~J.,  1992, \araa, 30, 575

\bibitem[\protect\citeauthoryear{{Coppin}, {Swinbank}, {Neri}, {Cox},
  {Alexander}, {Smail}, {Page}, {Stevens}, {Knudsen}, {Ivison}, {Beelen},
  {Bertoldi} \& {Omont}}{{Coppin} et~al.}{2008}]{Coppin:08}
{Coppin} K.~E.~K.,  {Swinbank} A.~M.,  {Neri} R.,  {Cox} P.,  {Alexander}
  D.~M.,  {Smail} I.,  {Page} M.~J.,  {Stevens} J.~A.,  {Knudsen} K.~K.,
  {Ivison} R.~J.,  {Beelen} A.,  {Bertoldi} F.,    {Omont} A.,  2008, \mnras,
  389, 45

\bibitem[\protect\citeauthoryear{{Danielson}, {Swinbank}, {Smail}, {Cox},
  {Edge}, {Weiss}, {Harris}, {Baker}, {De Breuck}, {Geach}, {Ivison}, {Krips},
  {Lundgren}, {Longmore}, {Neri} \& {Flaquer}}{{Danielson}
  et~al.}{2011}]{Danielson:11}
{Danielson} A.~L.~R.,  {Swinbank} A.~M.,  {Smail} I.,  {Cox} P.,  {Edge} A.~C.,
   {Weiss} A.,  {Harris} A.~I.,  {Baker} A.~J.,  {De Breuck} C.,  {Geach}
  J.~E.,  {Ivison} R.~J.,  {Krips} M.,  {Lundgren} A.,  {Longmore} S.,  {Neri}
  R.,    {Flaquer} B.~O.,  2011, \mnras, 410, 1687

\bibitem[\protect\citeauthoryear{{Draine} et~al.,}{{Draine}
  et~al.}{2007}]{Draine:07}
{Draine} B.~T.,  et~al., 2007, \apj, 663, 866

\bibitem[\protect\citeauthoryear{{Feroz} \& {Hobson}}{{Feroz} \&
  {Hobson}}{2008}]{Feroz:08}
{Feroz} F.,  {Hobson} M.~P.,  2008, \mnras, 384, 449

\bibitem[\protect\citeauthoryear{{Feruglio}, {Bongiorno}, {Fiore}, {Krips},
  {Brusa}, {Daddi}, {Gavignaud}, {Maiolino}, {Piconcelli}, {Sargent}, {Vignali}
  \& {Zappacosta}}{{Feruglio} et~al.}{2014}]{Feruglio:14}
{Feruglio} C.,  {Bongiorno} A.,  {Fiore} F.,  {Krips} M.,  {Brusa} M.,  {Daddi}
  E.,  {Gavignaud} I.,  {Maiolino} R.,  {Piconcelli} E.,  {Sargent} M.,
  {Vignali} C.,    {Zappacosta} L.,  2014, \aap, 565, A91

\bibitem[\protect\citeauthoryear{{Glover} \& {Clark}}{{Glover} \&
  {Clark}}{2016}]{Glover:16}
{Glover} S.~C.~O.,  {Clark} P.~C.,  2016, \mnras, 456, 3596

\bibitem[\protect\citeauthoryear{{Gullberg}, {Lehnert}, {De Breuck}, {Branchu},
  {Dannerbauer}, {Drouart}, {Emonts}, {Guillard}, {Hatch}, {Nesvadba}, {Omont},
  {Seymour} \& {Vernet}}{{Gullberg} et~al.}{2016}]{Gullberg:16}
{Gullberg} B.,  {Lehnert} M.~D.,  {De Breuck} C.,  {Branchu} S.,  {Dannerbauer}
  H.,  {Drouart} G.,  {Emonts} B.,  {Guillard} P.,  {Hatch} N.,  {Nesvadba}
  N.~P.~H.,  {Omont} A.,  {Seymour} N.,    {Vernet} J.,  2016, \aap, 591, A73

\bibitem[\protect\citeauthoryear{{Hopkins}, {Hernquist}, {Cox} \& {Kere{\v
  s}}}{{Hopkins} et~al.}{2008}]{Hopkins:08}
{Hopkins} P.~F.,  {Hernquist} L.,  {Cox} T.~J.,    {Kere{\v s}} D.,  2008,
  \apjs, 175, 356

\bibitem[\protect\citeauthoryear{{Ivison} et~al.,}{{Ivison}
  et~al.}{2012}]{Ivison:12}
{Ivison} R.~J.,  et~al., 2012, \mnras, 425, 1320

\bibitem[\protect\citeauthoryear{{Kormendy} \& {Ho}}{{Kormendy} \&
  {Ho}}{2013}]{Kormendy:13}
{Kormendy} J.,  {Ho} L.~C.,  2013, \araa, 51, 511

\bibitem[\protect\citeauthoryear{{Magorrian}, {Tremaine}, {Richstone},
  {Bender}, {Bower}, {Dressler}, {Faber}, {Gebhardt}, {Green}, {Grillmair},
  {Kormendy} \& {Lauer}}{{Magorrian} et~al.}{1998}]{Magorrian:98}
{Magorrian} J.,  {Tremaine} S.,  {Richstone} D.,  {Bender} R.,  {Bower} G.,
  {Dressler} A.,  {Faber} S.~M.,  {Gebhardt} K.,  {Green} R.,  {Grillmair} C.,
  {Kormendy} J.,    {Lauer} T.,  1998, \aj, 115, 2285

\bibitem[\protect\citeauthoryear{{Meijerink}, {Spaans} \& {Israel}}{{Meijerink}
  et~al.}{2006}]{Meijerink:06}
{Meijerink} R.,  {Spaans} M.,    {Israel} F.~P.,  2006, \apjl, 650, L103

\bibitem[\protect\citeauthoryear{{Meijerink}, {Spaans}, {Loenen} \& {van der
  Werf}}{{Meijerink} et~al.}{2011}]{Meijerink:11}
{Meijerink} R.,  {Spaans} M.,  {Loenen} A.~F.,    {van der Werf} P.~P.,  2011,
  \aap, 525, A119

\bibitem[\protect\citeauthoryear{{Momjian}, {Carilli}, {Walter} \&
  {Venemans}}{{Momjian} et~al.}{2014}]{Momjian:14}
{Momjian} E.,  {Carilli} C.~L.,  {Walter} F.,    {Venemans} B.,  2014, \aj,
  147, 6

\bibitem[\protect\citeauthoryear{{Omont}, {Cox}, {Bertoldi}, {McMahon},
  {Carilli} \& {Isaak}}{{Omont} et~al.}{2001}]{Omont:01}
{Omont} A.,  {Cox} P.,  {Bertoldi} F.,  {McMahon} R.~G.,  {Carilli} C.,
  {Isaak} K.~G.,  2001, \aap, 374, 371

\bibitem[\protect\citeauthoryear{{Papadopoulos}, {Thi} \&
  {Viti}}{{Papadopoulos} et~al.}{2004}]{Papadopoulos:04}
{Papadopoulos} P.~P.,  {Thi} W.-F.,    {Viti} S.,  2004, \mnras, 351, 147

\bibitem[\protect\citeauthoryear{{Papadopoulos}, {van der Werf}, {Xilouris},
  {Isaak}, {Gao} \& {M{\"u}hle}}{{Papadopoulos} et~al.}{2012}]{Papadopoulos:12}
{Papadopoulos} P.~P.,  {van der Werf} P.~P.,  {Xilouris} E.~M.,  {Isaak} K.~G.,
   {Gao} Y.,    {M{\"u}hle} S.,  2012, \mnras, 426, 2601

\bibitem[\protect\citeauthoryear{{Pelupessy}, {Papadopoulos} \& {van der
  Werf}}{{Pelupessy} et~al.}{2006}]{Pelupessy:06}
{Pelupessy} F.~I.,  {Papadopoulos} P.~P.,    {van der Werf} P.,  2006, \apj,
  645, 1024

\bibitem[\protect\citeauthoryear{{Priddey}, {Isaak}, {McMahon} \&
  {Omont}}{{Priddey} et~al.}{2003}]{Priddey:03}
{Priddey} R.~S.,  {Isaak} K.~G.,  {McMahon} R.~G.,    {Omont} A.,  2003,
  \mnras, 339, 1183

\bibitem[\protect\citeauthoryear{{R{\'e}my-Ruyer} et~al.,}{{R{\'e}my-Ruyer}
  et~al.}{2014}]{Remy-Ruyer:14}
{R{\'e}my-Ruyer} A.,  et~al., 2014, \aap, 563, A31

\bibitem[\protect\citeauthoryear{{Riechers} et~al.,}{{Riechers}
  et~al.}{2011a}]{Riechers:11b}
{Riechers} D.~A.,  et~al., 2011a, \apjl, 739, L32

\bibitem[\protect\citeauthoryear{{Riechers} et~al.,}{{Riechers}
  et~al.}{2011b}]{Riechers:11a}
{Riechers} D.~A.,  et~al., 2011b, \apjl, 739, L31

\bibitem[\protect\citeauthoryear{{Sanders}, {Soifer}, {Elias}, {Madore},
  {Matthews}, {Neugebauer} \& {Scoville}}{{Sanders} et~al.}{1988}]{Sanders:88}
{Sanders} D.~B.,  {Soifer} B.~T.,  {Elias} J.~H.,  {Madore} B.~F.,  {Matthews}
  K.,  {Neugebauer} G.,    {Scoville} N.~Z.,  1988, \apj, 325, 74

\bibitem[\protect\citeauthoryear{{Sandstrom} et~al.,}{{Sandstrom}
  et~al.}{2013}]{Sandstrom:13}
{Sandstrom} K.~M.,  et~al., 2013, \apj, 777, 5

\bibitem[\protect\citeauthoryear{{Santini} et~al.,}{{Santini}
  et~al.}{2014}]{Santini:14}
{Santini} P.,  et~al., 2014, \aap, 562, A30

\bibitem[\protect\citeauthoryear{{Scoville}, {Aussel}, {Sheth}, {Scott},
  {Sanders}, {Ivison}, {Pope}, {Capak}, {Vanden Bout}, {Manohar}, {Kartaltepe},
  {Robertson} \& {Lilly}}{{Scoville} et~al.}{2014}]{Scoville:14}
{Scoville} N.,  {Aussel} H.,  {Sheth} K.,  {Scott} K.~S.,  {Sanders} D.,
  {Ivison} R.,  {Pope} A.,  {Capak} P.,  {Vanden Bout} P.,  {Manohar} S.,
  {Kartaltepe} J.,  {Robertson} B.,    {Lilly} S.,  2014, \apj, 783, 84

\bibitem[\protect\citeauthoryear{{Sharon}, {Riechers}, {Hodge}, {Carilli},
  {Walter}, {Wei{\ss}}, {Knudsen} \& {Wagg}}{{Sharon} et~al.}{2016}]{Sharon:16}
{Sharon} C.~E.,  {Riechers} D.~A.,  {Hodge} J.,  {Carilli} C.~L.,  {Walter} F.,
   {Wei{\ss}} A.,  {Knudsen} K.~K.,    {Wagg} J.,  2016, \apj, 827, 18

\bibitem[\protect\citeauthoryear{{Takata}, {Sekiguchi}, {Smail}, {Chapman},
  {Geach}, {Swinbank}, {Blain} \& {Ivison}}{{Takata} et~al.}{2006}]{Takata:06}
{Takata} T.,  {Sekiguchi} K.,  {Smail} I.,  {Chapman} S.~C.,  {Geach} J.~E.,
  {Swinbank} A.~M.,  {Blain} A.,    {Ivison} R.~J.,  2006, \apj, 651, 713

\bibitem[\protect\citeauthoryear{{Venemans}, {Walter}, {Decarli}, {Ferkinhoff},
  {Weiss}, {Findlay}, {McMahon}, {Sutherland} \& {Meijerink}}{{Venemans}
  et~al.}{2017}]{Venemans:17}
{Venemans} B.,  {Walter} F.,  {Decarli} R.,  {Ferkinhoff} C.,  {Weiss} A.,
  {Findlay} J.,  {McMahon} R.,  {Sutherland} W.,    {Meijerink} R.,  2017,
  ArXiv e-prints

\bibitem[\protect\citeauthoryear{{Wagg}, {Carilli}, {Aravena}, {Cox},
  {Lentati}, {Maiolino}, {McMahon}, {Riechers}, {Walter}, {Andreani}, {Hills}
  \& {Wolfe}}{{Wagg} et~al.}{2014}]{Wagg:14}
{Wagg} J.,  {Carilli} C.~L.,  {Aravena} M.,  {Cox} P.,  {Lentati} L.,
  {Maiolino} R.,  {McMahon} R.~G.,  {Riechers} D.,  {Walter} F.,  {Andreani}
  P.,  {Hills} R.,    {Wolfe} A.,  2014, \apj, 783, 71

\bibitem[\protect\citeauthoryear{{Walter}, {Carilli}, {Bertoldi}, {Menten},
  {Cox}, {Lo}, {Fan} \& {Strauss}}{{Walter} et~al.}{2004}]{Walter:04}
{Walter} F.,  {Carilli} C.,  {Bertoldi} F.,  {Menten} K.,  {Cox} P.,  {Lo}
  K.~Y.,  {Fan} X.,    {Strauss} M.~A.,  2004, \apjl, 615, L17

\bibitem[\protect\citeauthoryear{{Walter}, {Wei{\ss}}, {Downes}, {Decarli} \&
  {Henkel}}{{Walter} et~al.}{2011}]{Walter:11}
{Walter} F.,  {Wei{\ss}} A.,  {Downes} D.,  {Decarli} R.,    {Henkel} C.,
  2011, \apj, 730, 18

\bibitem[\protect\citeauthoryear{{Wang} et~al.,}{{Wang}
  et~al.}{2016}]{Wang:16}
{Wang} R.,  et~al., 2016, \apj, 830, 53

\bibitem[\protect\citeauthoryear{{Wei{\ss}}, {Downes}, {Henkel} \&
  {Walter}}{{Wei{\ss}} et~al.}{2005}]{Weiss:05}
{Wei{\ss}} A.,  {Downes} D.,  {Henkel} C.,    {Walter} F.,  2005, \aap, 429,
  L25

\bibitem[\protect\citeauthoryear{{Wei{\ss}}, {Henkel}, {Downes} \&
  {Walter}}{{Wei{\ss}} et~al.}{2003}]{Weiss:03}
{Wei{\ss}} A.,  {Henkel} C.,  {Downes} D.,    {Walter} F.,  2003, \aap, 409,
  L41

\bibitem[\protect\citeauthoryear{{Willott}, {Omont} \& {Bergeron}}{{Willott}
  et~al.}{2013}]{Willott:13}
{Willott} C.~J.,  {Omont} A.,    {Bergeron} J.,  2013, \apj, 770, 13

\bibitem[\protect\citeauthoryear{{Yun} \& {Carilli}}{{Yun} \&
  {Carilli}}{2002}]{Yun:02}
{Yun} M.~S.,  {Carilli} C.~L.,  2002, \apj, 568, 88

\end{thebibliography}





\bsp	
\label{lastpage}
\end{document}